\documentclass[12pt,a4paper]{article}

\usepackage{tikz}

\usepackage{amsmath,amsfonts,amsthm,
graphicx,
lscape}

\usepackage[dvips]{psfrag}
\usepackage{cite}

\usepackage[normalem]{ulem}

\usepackage{textcomp}
\usepackage{amsmath,amsthm,amssymb,mathrsfs,cases}
\usepackage{amsfonts,graphicx,lscape}

\newcommand{\vt}[1]{\mbox{\boldmath$#1$}}

\numberwithin{equation}{section}

\def\beqa{\begin{eqnarray}}
\def\enqa{\end{eqnarray}}
\def\beq{\begin{equation}}
\def\enq{\end{equation}}

\begin{document}
\title{
On an 
integrable 
discretization
of 
the massive Thirring model in light-cone 
coordinates 
and the associated Yang--Baxter map 
}
\author{Takayuki \textsc{Tsuchida}
}
\maketitle
\begin{abstract} 
We propose 
a fully discrete analog of the massive Thirring model in light-cone 
coordinates 
by constructing its  
Lax-pair representation. 
This 
Lax-pair 
representation 
can also be used 
to define 
a new Yang--Baxter map, so we obtain a Yang--Baxter map 
that admits 
a continuous limit. 
We present 
most of 
the results 
for the general case where 
the dependent variables are 
matrix-valued. 
\end{abstract}
%

\newpage
\noindent
\tableofcontents

\newpage
\section{Introduction}

In the early 1990s, V.~G.~Drinfeld~\cite{Dri} 
suggested 
the problem 
of studying the set-theoretical solutions to 
the quantum Yang--Baxter equation:
\begin{equation}
R_{12} R_{13} R_{23} =R_{23} R_{13} R_{12}, 
\label{YB_equation}
\end{equation}
where 
$R_{ij}$ 
acts as a map \mbox{$X \times X \to X\times X$} 
on the  $i$th and $j$th factors 
of the 
Cartesian
product \mbox{$X \times X \times X$} 
and identically on the remaining factor, 
and $X$ is a set. 

Veselov~\cite{Ve03,Ve07} suggested to use 
the shorter term 
`Yang--Baxter map' instead of 
`set-theoretical solution to 
the quantum Yang--Baxter equation', which is now 
more common, 
so we also adopt it here. 
Since the suggestion by Drinfeld~\cite{Dri}, 
the study of Yang--Baxter maps has attracted much attention 
and 
now the number of known Yang--Baxter maps is 
enormous 
and 
the results on Yang--Baxter maps 
appear to be 
somewhat scattered, except for some 
systematic classification results (see, {\em e.g.}, \cite{ABS04}). 
This 
implies that the original 
problem setting 
by Drinfeld~\cite{Dri} 
would be 
too general, 
depending on the nature of the set
$X$, and there exist too many 
(sometimes 
infinitely many) 
solutions to the quantum Yang--Baxter equation (\ref{YB_equation}). 
Thus, 
it would be more realistic 
and 
meaningful 
to consider 
narrower versions 
of the Drinfeld problem,  
{\em e.g.}, by restricting 
the admissible form of the map $R$ 
in a natural manner. 

In this paper, we 
aim to construct a
Yang--Baxter map 
that admits a natural continuous limit. 
In the simplest case, 
such a map $R$ 
involves 
an arbitrary 
(typically small) 
parameter $\hbar$ 
and takes the following form~\cite{Skly82}: 
\begin{equation}
R= I + \hbar \hspace{1pt} r + \mathcal{O}(\hbar^2), 
\label{quasi_c}
\end{equation}
where $I$ is the identity mapping. 
The asymptotic expansion (\ref{quasi_c}) with respect to 
$\hbar$ 
is 
a natural ansatz 
in that 
the  quantum Yang--Baxter equation (\ref{YB_equation}) 
for a linear operator $R$ 
reduces 
in the quasi-classical limit \mbox{$\hbar \to 0$} 
to the classical Yang--Baxter equation~\cite{BelaDrin82,Semenov}: 
\begin{equation}
[r_{12}, r_{13}] +[r_{12}, r_{23}] + [r_{13}, r_{23}]=0, 
\nonumber
\end{equation}
where $[ \;\, , \;\, ]$ is the commutator. 
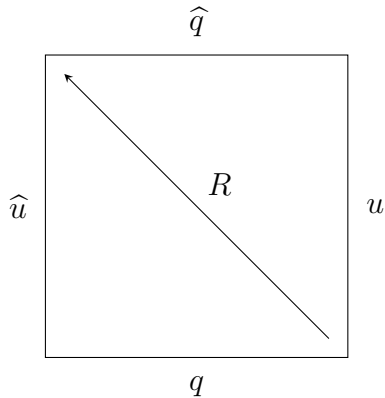
\begin{center}
\begin{tikzpicture}
\draw (0,0) rectangle (4,4);
\draw (2,-0.1)node[below]{$q$}; 
\draw (2,4.1)node[above] {$\widehat{q}$} ; 
\draw (-0.1,2)node[left] {$\widehat{u}$} ; 
\draw (4.1,2)node[right] {$u$} ; 
\draw[->,>=stealth] (3.75,0.25)--(0.25,3.75);
\draw (2,2) node[above right] {$R$} ;
\end{tikzpicture}
\end{center}

More specifically, if $q, u \in X$ and 
\begin{equation}
R: (q, u) \to \left( \widehat{q}, \widehat{u} \right) := (q, u) 
+ \hbar \hspace{1pt} \left( f(q,u), g(q, u) \right) + \mathcal{O}(\hbar^2), 
\nonumber
\end{equation}
the map $R$ 
provides 
in the continuous limit \mbox{$\hbar \to 0$} 
a continuous system in \mbox{$1+1$} space-time dimensions: 
\begin{equation} 
\nonumber
\left\{ 
\begin{split}
& q_{t} = f(q,u), 
\\
& u_x = -g(q,u), 
\end{split} 
\right. 
\end{equation}
where the subscripts $t$ and $x$ denote
the 
partial 
differentiation 
with respect to 
these variables. 
In our previous paper~\cite{Me_YB}, 
we already constructed a 
Yang--Baxter map 
of this type 
(see also 
some relevant results in~\cite{Terng98,Kouloukas2011,DM2019,DM2020,Caudrelier22,Caudrelier23}). 

We can consider a more general situation where 
the map $R_{12}$ 
admits the expansion in terms of (typically small) parameters $h$ and $\varDelta$  
as  
\begin{equation}
R_{12}: (q, u) \to \left( \widehat{q}, \widehat{u} \right) := (q, u) 
+ \left( h f(q,u), \varDelta g(q, u) \right) + (\mathcal{O}(h^2, h \varDelta),  \mathcal{O}(h \varDelta, \varDelta^2)). 
\label{quasi_c2}
\end{equation}
In this paper, we propose 
a Yang--Baxter map 
that admits the expansion 
(\ref{quasi_c2}) 
and provides in the continuous limit \mbox{$h \to 0$} or/and \mbox{$\varDelta \to 0$} 
a nontrivial semi-discrete/continuous system in \mbox{$1+1$} space-time dimensions. 
The Yang--Baxter map proposed in this paper 
appears to be 
more general and interesting 
than 
the Yang--Baxter map constructed 
in our previous paper~\cite{Me_YB}; 
closely related results were obtained previously 
by Xu and 
Pelinovsky~\cite{Pelinovsky2019}, but 
they did not explicitly 
present their results in the form of a Yang--Baxter map. 

This paper is organized as follows. 
In section~2, as the main building block of our construction, 
we introduce a discrete Lax 
matrix, 
which is motivated by 
a binary  
B\"acklund--Darboux transformation~\cite{Holod78,
Nissimov79,Morris80, Pri81,David84} 
for the continuous 
derivative nonlinear Schr\"odinger (Chen--Lee--Liu~\cite{CLL}) hierarchy. 
By solving the discrete zero-curvature condition 
based on 
this Lax matrix, 
we obtain a fully discrete analog of the massive Thirring model in light-cone 
(or characteristic) 
coordinates. 
We discuss how this discrete massive Thirring model reduces to the continuous 
massive Thirring model in 
a suitable 
continuous limit. 
In section~3, 
we present a binary 
B\"acklund--Darboux transformation 
for the 
discrete massive Thirring model, which can be used to construct its explicit solutions. 
In section~4, we present some symmetry properties of our 
discrete 
Lax matrix 
and, 
using these properties, we translate 
the results in section~2 
to the form of 
a 
Yang--Baxter map that admits a continuous limit. 
Section~5 is devoted to concluding remarks.

\section{Massive 
Thirring model}

\subsection{Continuous model}

The Lax-pair representation for 
the continuous massive Thirring model in light-cone 
coordinates~\cite{KN2,Morris79,GIK} 
(or its matrix generalization~\cite{Linden1,TW3}) 
is given by 
\begin{align}
& \mathrm{i}
\left[
\begin{array}{c}
 \Psi_{1} \\
 \Psi_{2} \\
\end{array}
\right]_{x}
= 
\left[
\begin{array}{cc}
 -q r & \zeta  q \\
 \zeta r & - \zeta^2 I \\
\end{array}
\right]
\left[
\begin{array}{c}
 \Psi_{1} \\
 \Psi_{2} \\
\end{array}
\right],
\label{AL-Mn1}
\end{align}
and
\begin{align}
& \mathrm{i} \left[
\begin{array}{c}
 \Psi_{1} \\
 \Psi_{2} \\
\end{array}
\right]_{t}
= 
\left[
\begin{array}{cc}
 -\frac{1}{\zeta^2} I & \frac{1}{\zeta} u \\[1mm]
 \frac{1}{\zeta} v & 
  - v u \\
\end{array}
\right]
\left[
\begin{array}{c}
 \Psi_{1} \\
 \Psi_{2} \\
\end{array}
\right].
\label{AL-Mn2}
\end{align}
Here, $\zeta$ is a constant spectral parameter and 
the subscripts $x$ and $t$ 
denote
the 
partial 
differentiation 
with respect to 
these variables. 
The Lax-pair representation for 
the 
massive Thirring model in laboratory  
coordinates~\cite{Kuz} can be obtained by applying a linear 
transformation 
of the independent variables. 
In this paper, 
unless otherwise specified, 
every 
square 
matrix is 
partitioned into four blocks as an 
\mbox{$(M+N) \times (M+N)$} matrix; 
we omit the subscripts of the identity matrix $I$ and the zero matrix $O$ 
to denote their matrix dimensions. 

The compatibility condition 
for the overdetermined linear equations 
(\ref{AL-Mn1}) and (\ref{AL-Mn2})
provides 
the equations of motion for 
the massive Thirring model in light-cone 
coordinates~\cite{KN2,Morris79,GIK} 
(see \cite{Linden1,TW3} for the matrix generalization): 
\begin{equation} 
\label{sd-AL1}
\left\{ 
\begin{split}
& \mathrm{i} q_{t}+ u - q v u =O, 
\\[0.5mm]
& \mathrm{i} r_{t}- v + v u r =O,
\\[0.5mm]
& \mathrm{i} u_{x}- q + q r u =O, 
\\[0.5mm]
& \mathrm{i} v_{x}+ r - v q r =O.
\end{split} 
\right. 
\end{equation}
%
Here and hereafter, we use the symbol italic $O$, 
instead of $0$, 
on the right-hand sides of the equations 
to 
imply 
that 
the dependent variables 
appearing on the left-hand sides 
can be 
matrix-valued. 

In the case of the scalar dependent variables, 
by imposing the complex conjugation reduction
\begin{equation}
r= q^\ast, \hspace{5mm} v=u^\ast, 
\nonumber
\end{equation}
where the asterisk denotes the complex conjugate, 
we obtain a 
familiar form of the massive Thirring model in light-cone 
coordinates: 
\begin{equation} 
\label{cMTM}
\left\{ 
\begin{split}
& \mathrm{i} q_{t}+ u - |u|^2q=0, 
\\[0.5mm]
& \mathrm{i} u_{x}- q + |q|^2 u =0.
\end{split} 
\right. 
\end{equation}

In the case of the matrix dependent variables, 
by imposing the Hermitian conjugation reduction
\begin{equation}
r= q^\dagger, \hspace{5mm} v=u^\dagger, 
\nonumber
\end{equation}
where the dagger denotes the Hermitian conjugate, 
we obtain the matrix generalization of (\ref{cMTM}): 
\begin{equation} 
\label{mMTM}
\left\{ 
\begin{split}
& \mathrm{i} q_{t}+ u - q u^\dagger u =O, 
\\[0.5mm]
& \mathrm{i} u_{x}- q + q q^\dagger u =O.
\end{split} 
\right. 
\end{equation}

The spectral problem (\ref{AL-Mn1}) 
is 
associated with 
the 
continuous derivative nonlinear Schr\"odinger 
(known as the Chen--Lee--Liu~\cite{CLL}) hierarchy~\cite{WS,Dodd83,Dodd84} 
(see~\cite{Linden1,TW3} for the matrix case);
the massive Thirring model in light-cone 
coordinates can be considered as 
the first negative flow of 
the continuous 
Chen--Lee--Liu 
hierarchy~\cite{NCQL,Linden1}. 

In the remaining part of this section, 
we consider how the 
above 
Lax-pair representation 
for the massive Thirring model in light-cone 
coordinates can be fully discretized.

\subsection{Discrete Lax pair}
\label{subsec2.2}

A proper discretization 
of the Chen--Lee--Liu 
spectral problem (\ref{AL-Mn1}) 
can be 
obtained from a 
suitable 
B\"acklund--Darboux transformation 
for the continuous 
Chen--Lee--Liu hierarchy~\cite{Me2015}. 
Specifically, 
we
consider 
the following discrete spectral 
problem: 
\begin{equation}
\left[
\begin{array}{c}
 \Psi_{1, n+1}  \\
 \Psi_{2, n+1} \\
\end{array}
\right]
= {\cal L}  (q_n, r_n; \mu, \nu, \xi, \eta)  
\left[
\begin{array}{c}
 \Psi_{1,n}  \\
 \Psi_{2,n} \\
\end{array}
\right], 
\label{gDB2}
\end{equation}
where \mbox{$n \in \mathbb{Z}$} represents the discrete spatial coordinate and 
the discrete spatial 
Lax matrix ${\cal L} (q_n, r_n; \mu, \nu, \xi, \eta)$ 
is defined as 
\begin{align}
& 
{\cal L} (q_n, r_n; \mu, \nu, \xi, \eta) 
\nonumber \\[2mm]
&:=  \left[
\begin{array}{cc}
 \mu I & O \\ 
 O & \xi I \\
\end{array}
\right] + \frac{\mu \nu - \xi \eta}{\xi \zeta^2 + \nu}
 \left[
\begin{array}{cc}
 q_n r_n (\xi \eta I - q_n r_n )^{-1} &  -\xi \zeta q_n (\xi \eta I - r_n q_n)^{-1}  \\
 -\xi \zeta r_n (\xi \eta I - q_n r_n )^{-1} &  \xi^2 \zeta^2 (\xi \eta I - r_n q_n)^{-1}  \\
\end{array}
\right]
\nonumber \\[2mm]
&\hphantom{:}=  \left[
\begin{array}{cc}
 \mu I & O \\ 
 O & \xi I \\
\end{array}
\right] + \frac{\mu \nu - \xi \eta}{\xi \zeta^2 + \nu}
 \left[
\begin{array}{c}
 q_n \\
 -\xi \zeta I \\
\end{array}
\right]
(\xi \eta I - r_n q_n)^{-1}  
\left[
\begin{array}{cc}
 r_n & - \xi \zeta I \\
\end{array}
\right]. 
\label{Ln_YB}
\end{align}
Here, $\zeta$ is the spectral parameter common to all the Lax matrices 
appearing in this paper, 
so the dependence of ${\cal L}$ on $\zeta$ is suppressed;  
$\mu$, $\nu$, $\xi$ and $\eta$
are 
nonzero 
constant parameters 
satisfying the condition
\mbox{$\mu \nu - \xi \eta \neq 0$}. 
%
%
%
The matrix (\ref{Ln_YB})
is equivalent, up to an 
unessential overall factor, 
to the spatial Lax matrix considered in our previous paper~\cite{Me2015}. 


We use the tilde $\widetilde{}$ 
to denote the forward 
shift (\mbox{$m \to m+1$})
in the discrete time coordinate \mbox{$m \in {\mathbb Z}$}~\cite{Suris03} 
and consider the following discrete-time evolution: 
\begin{equation}
\left[
\begin{array}{c}
 \widetilde{\Psi}_{1, n}  \\
 \widetilde{\Psi}_{2, n} \\
\end{array}
\right]
= {\cal L}(u_n, v_n; \alpha, \beta, \gamma, \delta)
\left[
\begin{array}{c}
 \Psi_{1,n}  \\
 \Psi_{2,n} \\
\end{array}
\right],
\label{gDB_time}
\end{equation}
where 
the discrete temporal 
Lax matrix ${\cal L}(u_n, v_n; \alpha, \beta, \gamma, \delta)$  
is defined 
exactly in the same manner 
as 
in 
(\ref{Ln_YB}): 
\begin{align}
& 
{\cal L} (u_n, v_n; \alpha, \beta, \gamma, \delta) 
\nonumber \\[2mm]
&:=  \left[
\begin{array}{cc}
 \alpha I & O \\ 
 O & \gamma I \\
\end{array}
\right] + \frac{\alpha \beta - \gamma \delta}{\gamma \zeta^2 + \beta}
 \left[
\begin{array}{cc}
 u_n v_n (\gamma \delta I - u_n v_n )^{-1} &  -\gamma \zeta u_n (\gamma \delta I - v_n u_n)^{-1}  \\
 -\gamma \zeta v_n (\gamma \delta I - u_n v_n )^{-1} &  \gamma^2 \zeta^2 (\gamma \delta I - v_n u_n)^{-1}  \\
\end{array}
\right]
\nonumber \\[2mm]
&\hphantom{:}=  \left[
\begin{array}{cc}
 \alpha I & O \\ 
 O & \gamma I \\
\end{array}
\right] + \frac{\alpha \beta - \gamma \delta}{\gamma \zeta^2 + \beta}
 \left[
\begin{array}{c}
 u_n \\
 -\gamma \zeta I \\
\end{array}
\right]
(\gamma \delta I - v_n u_n)^{-1}  
\left[
\begin{array}{cc}
 v_n & - \gamma \zeta I \\
\end{array}
\right]. 
\label{Ln_YB2}
\end{align}
Here, $\alpha$, $\beta$, $\gamma$ and $\delta$
are 
nonzero constant parameters 
satisfying the condition 
\mbox{$\alpha \beta - \gamma \delta \neq 0$}. 

The compatibility condition for 
the overdetermined 
linear 
equations (\ref{gDB2}) and (\ref{gDB_time}) 
is given by~\cite{AL76,AL77,Orfa1}
%
\begin{equation}
 {\cal L}  (\widetilde{q}_n, \widetilde{r}_n; \mu, \nu, \xi, \eta)  {\cal L}(u_n, v_n; \alpha, \beta, \gamma, \delta)
 =  {\cal L}(u_{n+1}, v_{n+1}; \alpha, \beta, \gamma, \delta) {\cal L}  (q_n, r_n; \mu, \nu, \xi, \eta), 
\label{fd-Lax}
\end{equation}
which is 
a fully discrete analog of 
the zero-curvature 
condition. 
Hereafter, 
we assume that
the 
nonzero 
parameters 
$\mu$, $\nu$, $\xi$, $\eta$, $\alpha$, $\beta$, $\gamma$ and $\delta$
satisfy 
the conditions 
\mbox{$\alpha \nu - \delta \xi \neq 0$}, \mbox{$\alpha \eta - \delta \mu \neq 0$}, 
\mbox{$\beta \mu - \gamma \eta \neq 0$} and \mbox{$\beta \xi - \gamma \nu \neq 0$}, 
in addition to the conditions \mbox{$\mu \nu - \xi \eta \neq 0$} and \mbox{$\alpha \beta - \gamma \delta \neq 0$}. 
In a nutshell, 
the values of $\nu/\xi$, $\eta/\mu$, $\beta/\gamma$ and $\delta/\alpha$ 
are 
assumed to be 
all different. 

\begin{center}
\begin{tikzpicture}
\node (A) at (0,0) {};
\node (B) at (4,0) {}; 
\node (C) at (0,4) {}; 
\node (D) at (4,4) {}; 
\fill (A) circle (2pt) (B) circle (2pt) (C) circle (2pt) (D) circle (2pt); 
\draw[->] (A)--(B);
\draw[->] (A)--(C);
\draw[->] (C)--(D);
\draw[->] (B)--(D);
\draw (2,-0.1)node[below]{${\cal L}  (q_n, r_n; \mu, \nu, \xi, \eta)$}; 
\draw (2,4.1)node[above] {${\cal L}  (\widetilde{q}_n, \widetilde{r}_n; \mu, \nu, \xi, \eta)$} ; 
\draw (-0.1,2)node[left] {${\cal L}(u_n, v_n; \alpha, \beta, \gamma, \delta)$} ; 
\draw (4.1,2)node[right] {${\cal L}(u_{n+1}, v_{n+1}; \alpha, \beta, \gamma, \delta)$} ; 
\draw (0,0)node[below left]
{$\left[
\begin{array}{c}
 \Psi_{1,n} \\
 \Psi_{2,n} \\
\end{array}
\right]$}; 
\draw (0,4)node[above left]
{$\left[
\begin{array}{c}
 \widetilde{\Psi}_{1,n} \\
 \widetilde{\Psi}_{2,n} \\
\end{array}
\right]$}; 
\draw (4,0)node[below right]
{$\left[
\begin{array}{c}
 \Psi_{1,n+1} \\
 \Psi_{2,n+1} \\
\end{array}
\right]$}; 
\draw (4,4)node[above right]
{$\left[
\begin{array}{c}
 \widetilde{\Psi}_{1,n+1} \\
 \widetilde{\Psi}_{2,n+1} \\
\end{array}
\right]$}; 
\end{tikzpicture}
\end{center}

Clearly, we can multiply the discrete 
spatial/temporal Lax matrix 
by 
any nonzero constant factor without changing 
the discrete zero-curvature 
condition (\ref{fd-Lax}). 
Alternatively but equivalently, 
we can consider the gauge transformation: 
\begin{equation}
\left[
\begin{array}{c}
 \Psi_{1, n}  \\
 \Psi_{2, n} \\
\end{array}
\right]
\to a^n b^m 
\left[
\begin{array}{c}
 \Psi_{1,n}  \\
 \Psi_{2,n} \\
\end{array}
\right], 
\nonumber 
\end{equation}
where $m$ is the discrete time coordinate, and 
$a$ and $b$ are arbitrary nonzero constant factors that may depend on the spectral 
parameter $\zeta$. 
It is 
often more convenient to 
adopt the temporal Lax matrix of the following rescaled form:
\begin{align}
& 
\frac{\zeta + \frac{\beta}{\gamma \zeta}}{\frac{\alpha}{\delta} \zeta + \frac{1}{\zeta}}
{\cal L} (u_n, v_n; \alpha, \beta, \gamma, \delta) 
\nonumber \\[2mm]
&=  \left[
\begin{array}{cc}
 \delta I & O \\ 
 O & \beta I \\
\end{array}
\right] + \frac{\alpha \beta - \gamma \delta}{\alpha + \frac{\delta}{\zeta^2}}
 \left[
\begin{array}{cc}
 \frac{\delta^2}{\zeta^2} (\gamma \delta I - u_n v_n )^{-1} &  -\frac{\delta}{\zeta} u_n (\gamma \delta I - v_n u_n)^{-1}  \\[1mm]
 -\frac{\delta}{\zeta} v_n (\gamma \delta I - u_n v_n )^{-1} &  v_n u_n (\gamma \delta I - v_n u_n)^{-1}  \\
\end{array}
\right]
\nonumber \\[2mm]
&=  \left[
\begin{array}{cc}
 \delta I & O \\ 
 O & \beta I \\
\end{array}
\right] + \frac{\alpha \beta - \gamma \delta}{\alpha + \frac{\delta}{\zeta^2}}
 \left[
\begin{array}{c}
 -\frac{\delta}{\zeta} I \\
 v_n \\
\end{array}
\right]
(\gamma \delta I - u_n v_n)^{-1}  
\left[
\begin{array}{cc}
 -\frac{\delta}{\zeta} I & u_n \\
\end{array}
\right]. 
\label{Ln_YB3}
\end{align}

We 
note 
that 
\begin{align}
& \hphantom{:=}  
{\cal L} (q_n, r_n; 1, \nu/\varDelta, 1, \eta/\varDelta) 
\nonumber \\[2mm]
&=  I + \varDelta \frac{\nu - \eta}{\varDelta \zeta^2 + \nu}
 \left[
\begin{array}{cc}
 q_n r_n (\eta I - \varDelta q_n r_n )^{-1} &  -\zeta q_n (\eta I - \varDelta r_n q_n)^{-1}  \\
 -\zeta r_n (\eta I - \varDelta q_n r_n )^{-1} &  \zeta^2 (\eta I - \varDelta r_n q_n)^{-1}  \\
\end{array}
\right]
\nonumber \\[2mm]
&= I + \varDelta \frac{\nu - \eta}{\nu \eta}
 \left[
\begin{array}{cc}
 q_n r_n &  -\zeta q_n \\
 -\zeta r_n &  \zeta^2 I \\
\end{array}
\right] + \mathcal{O}(\varDelta^2), 
\nonumber
\end{align}
and
\begin{align}
& \hphantom{=}  
\frac{\zeta + \frac{h}{\gamma \zeta}}{\frac{\alpha}{h} \zeta + \frac{1}{\zeta}}
{\cal L} (u_n, v_n; \alpha/h, 1, \gamma/h, 1) 
\nonumber \\[2mm]
&=  I + h \frac{\alpha - \gamma}{\alpha+ \frac{h}{\zeta^2}}
 \left[
\begin{array}{cc}
 \frac{1}{\zeta^2} (\gamma I - h u_n v_n )^{-1} 
	&  -\frac{1}{\zeta} u_n (\gamma I - h v_n u_n)^{-1}  \\[1mm]
 -\frac{1}{\zeta} v_n (\gamma I - h u_n v_n )^{-1} &  v_n u_n ( \gamma I - h v_n u_n)^{-1}  \\
\end{array}
\right]
\nonumber \\[2mm]
&= I + h \frac{\alpha- \gamma}{\alpha \gamma}
 \left[
\begin{array}{cc}
 \frac{1}{\zeta^2} I &  -\frac{1}{\zeta} u_n \\[1mm]
 -\frac{1}{\zeta} v_n & v_n u_n  \\
\end{array}
\right] + \mathcal{O}(h^2). 
\nonumber
\end{align}
Thus, 
we can 
recover 
the spatial 
Lax 
matrix 
in 
(\ref{AL-Mn1}) 
by setting 
\begin{equation}
 \frac{\nu - \eta}{\nu \eta} 
= \mathrm{i},
\nonumber
\end{equation}
and taking the continuous space limit 
\mbox{$\varDelta \to 0$}; 
the temporal 
Lax 
matrix 
in 
(\ref{AL-Mn2}) 
can be recovered 
by setting 
\begin{equation}
 \frac{\alpha- \gamma}{\alpha \gamma} = \mathrm{i},
\nonumber
\end{equation}
and taking the continuous time limit 
\mbox{$h \to 0$}.

\subsection{Fully discrete equations of motion}
\label{subsec2.3}

Substituting (\ref{Ln_YB}) and (\ref{Ln_YB2}) 
into the discrete zero-curvature condition 
(\ref{fd-Lax}), 
we obtain 
\begin{align}
& 
 \left\{ \left[
\begin{array}{cc}
 \mu I & O \\ 
 O & \xi I \\
\end{array}
\right] + \frac{\mu \nu - \xi \eta}{\xi \zeta^2 + \nu}
 \left[
\begin{array}{c}
 \widetilde{q}_n \\
 -\xi \zeta I \\
\end{array}
\right]
(\xi \eta I - \widetilde{r}_n \widetilde{q}_n)^{-1}  
\left[
\begin{array}{cc}
 \widetilde{r}_n & - \xi \zeta I \\
\end{array}
\right] \right\} 
\nonumber \\[1mm] 
& \times \left\{ \left[
\begin{array}{cc}
 \alpha I & O \\ 
 O & \gamma I \\
\end{array}
\right] + \frac{\alpha \beta - \gamma \delta}{\gamma \zeta^2 + \beta}
 \left[
\begin{array}{c}
 u_n \\
 -\gamma \zeta I \\
\end{array}
\right]
(\gamma \delta I - v_n u_n)^{-1}  
\left[
\begin{array}{cc}
 v_n & - \gamma \zeta I \\
\end{array}
\right] \right\} 
\nonumber \\[2mm]
= & \left\{ \left[
\begin{array}{cc}
 \alpha I & O \\ 
 O & \gamma I \\
\end{array}
\right] + \frac{\alpha \beta - \gamma \delta}{\gamma \zeta^2 + \beta}
 \left[
\begin{array}{c}
 u_{n+1} \\
 -\gamma \zeta I \\
\end{array}
\right]
(\gamma \delta I - v_{n+1} u_{n+1})^{-1}  
\left[
\begin{array}{cc}
 v_{n+1} & - \gamma \zeta I \\
\end{array}
\right] \right\} 
\nonumber \\[1mm]
& \times \left\{ \left[
\begin{array}{cc}
 \mu I & O \\ 
 O & \xi I \\
\end{array}
\right] + \frac{\mu \nu - \xi \eta}{\xi \zeta^2 + \nu}
 \left[
\begin{array}{c}
 q_n \\
 -\xi \zeta I \\
\end{array}
\right]
(\xi \eta I - r_n q_n)^{-1}  
\left[
\begin{array}{cc}
 r_n & - \xi \zeta I \\
\end{array}
\right] \right\}. 
\label{LL=LL}
\end{align}

We subtract the common term:
\[
 \left[
\begin{array}{cc}
 \alpha \mu I & O \\ 
 O & \gamma \xi I \\
\end{array}
\right]
\] 
from both sides of 
(\ref{LL=LL}) and multiply 
it 
both from the left and 
right
by the 
diagonal matrix: 
\[
\left[
\begin{array}{cc}
 I & O \\ 
 O & \frac{1}{\zeta} I \\
\end{array}
\right]. 
\]
Then, 
because of the condition 
\mbox{$\beta \xi - \gamma \nu \neq 0$} 
(or, equivalently, \mbox{$\nu/\xi \neq \beta/\gamma$}) 
and 
the identities: 
\begin{align}
& \frac{1}{(\xi \zeta^2 + \nu)(\gamma \zeta^2 + \beta)} 
= \frac{1}{\beta \xi - \gamma \nu} \left( \frac{\xi}{\xi \zeta^2 + \nu} -  \frac{\gamma}{\gamma \zeta^2 + \beta} \right),
\nonumber \\[2mm]
& \frac{\zeta^2}{(\xi \zeta^2 + \nu)(\gamma \zeta^2 + \beta)} 
= \frac{1}{\beta \xi - \gamma \nu} \left( -\frac{\nu}{\xi \zeta^2 + \nu} + \frac{\beta}{\gamma \zeta^2 + \beta} \right), 
\nonumber 
\end{align}
we obtain from the coefficients of 
$1/(\xi \zeta^2 + \nu)$ and $1/(\gamma \zeta^2 + \beta)$ 
the following two relations: 
\begin{align}
& 
 \left[
\begin{array}{c}
 \widetilde{q}_n \\
 -\xi I \\
\end{array}
\right]
(\xi \eta I - \widetilde{r}_n \widetilde{q}_n)^{-1}  
\left[
\begin{array}{cc}
 \widetilde{r}_n & - \xi I \\
\end{array}
\right]
\nonumber \\[1mm]
& \times 
 \left\{ \left[
\begin{array}{cc}
 \alpha I & O \\ 
 O & \gamma I \\
\end{array}
\right] + \frac{\alpha \beta - \gamma \delta}{\beta \xi - \gamma \nu}
 \left[
\begin{array}{cc}
 \xi I & O \\ 
 O & -\nu I \\
\end{array}
\right]
 \left[
\begin{array}{c}
 u_n \\
 -\gamma I \\
\end{array}
\right]
(\gamma \delta I - v_n u_n)^{-1}  
\left[
\begin{array}{cc}
 v_n & - \gamma I \\
\end{array}
\right] \right\} 
\nonumber \\[2mm]
& = 
 \left\{ \left[
\begin{array}{cc}
 \alpha I & O \\ 
 O & \gamma I \\
\end{array}
\right] + \frac{\alpha \beta - \gamma \delta}{\beta \xi - \gamma \nu}
 \left[
\begin{array}{c}
 u_{n+1} \\
 -\gamma I \\
\end{array}
\right]
(\gamma \delta I - v_{n+1} u_{n+1})^{-1}  
\left[
\begin{array}{cc}
 v_{n+1} & - \gamma I \\
\end{array}
\right] 
 \left[
\begin{array}{cc}
 \xi I & O \\ 
 O & -\nu I \\
\end{array}
\right]
\right\} 
\nonumber \\[1mm]
& \times 
 \left[
\begin{array}{c}
 q_n \\
 -\xi I \\
\end{array}
\right]
(\xi \eta I - r_n q_n)^{-1}  
\left[
\begin{array}{cc}
 r_n & - \xi I \\
\end{array}
\right],
\label{q_r_tilde1} 
\end{align}
and 
\begin{align}
& \left\{ \left[
\begin{array}{cc}
 \mu I & O \\ 
 O & \xi I \\
\end{array}
\right] + \frac{\mu \nu - \xi \eta}{\beta \xi - \gamma \nu}
 \left[
\begin{array}{c}
 \widetilde{q}_{n} \\
 -\xi I \\
\end{array}
\right]
(\xi \eta I - \widetilde{r}_{n} \widetilde{q}_{n})^{-1}  
\left[
\begin{array}{cc}
 \widetilde{r}_{n} & - \xi I \\
\end{array}
\right] 
 \left[
\begin{array}{cc}
 -\gamma I & O \\ 
 O & \beta I \\
\end{array}
\right]
\right\} 
\nonumber \\[1mm]
& \times 
 \left[
\begin{array}{c}
 u_n \\
 -\gamma I \\
\end{array}
\right]
(\gamma \delta I - v_n u_n)^{-1}  
\left[
\begin{array}{cc}
 v_n & - \gamma I \\
\end{array}
\right]
\nonumber \\[2mm]
& = 
 \left[
\begin{array}{c}
 u_{n+1} \\
 -\gamma I \\
\end{array}
\right]
(\gamma \delta I - v_{n+1} u_{n+1})^{-1}  
\left[
\begin{array}{cc}
 v_{n+1} & - \gamma I \\
\end{array}
\right]
\nonumber \\[1mm]
& \times 
 \left\{ \left[
\begin{array}{cc}
 \mu I & O \\ 
 O & \xi I \\
\end{array}
\right] + \frac{\mu \nu - \xi \eta}{\beta \xi - \gamma \nu}
 \left[
\begin{array}{cc}
 -\gamma I & O \\ 
 O & \beta I \\
\end{array}
\right]
 \left[
\begin{array}{c}
 q_n \\
 -\xi I \\
\end{array}
\right]
(\xi \eta I - r_n q_n)^{-1}  
\left[
\begin{array}{cc}
 r_n & - \xi I \\
\end{array}
\right] \right\}. 
\label{u_v_1} 
\end{align}
Noting the identities: 
\begin{align}
& \left\{ \left[
\begin{array}{cc}
 \alpha I & O \\ 
 O & \gamma I \\
\end{array}
\right] + \frac{\alpha \beta - \gamma \delta}{\beta \xi - \gamma \nu}
 \left[
\begin{array}{cc}
 \xi I & O \\ 
 O & -\nu I \\
\end{array}
\right]
 \left[
\begin{array}{c}
 u_n \\
 -\gamma I \\
\end{array}
\right]
(\gamma \delta I - v_n u_n)^{-1}  
\left[
\begin{array}{cc}
 v_n & - \gamma I \\
\end{array}
\right] \right\} 
\nonumber \\[1mm]
&  \times 
\left\{ \left[
\begin{array}{cc}
 \gamma I & O \\ 
 O & \alpha I \\
\end{array}
\right] + \frac{\alpha \beta - \gamma \delta}{\alpha \nu - \delta \xi}
 \left[
\begin{array}{cc}
 \xi I & O \\ 
 O & -\nu I \\
\end{array}
\right]
 \left[
\begin{array}{c}
 u_n \\
 -\alpha I \\
\end{array}
\right]
(\alpha \beta I - v_n u_n)^{-1}  
\left[
\begin{array}{cc}
 v_n & - \alpha I \\
\end{array}
\right] \right\} 
\nonumber \\[1mm]
& =  \alpha \gamma I, 
\label{sharp1}
\end{align}
and 
\begin{align}
& \left\{ \left[
\begin{array}{cc}
 \mu I & O \\ 
 O & \xi I \\
\end{array}
\right] + \frac{\mu \nu - \xi \eta}{\beta \xi - \gamma \nu}
 \left[
\begin{array}{cc}
 -\gamma I & O \\ 
 O & \beta I \\
\end{array}
\right]
 \left[
\begin{array}{c}
 q_n \\
 -\xi I \\
\end{array}
\right]
(\xi \eta I - r_n q_n)^{-1}  
\left[
\begin{array}{cc}
 r_n & - \xi I \\
\end{array}
\right] \right\}
\nonumber \\[1mm]
& \times 
 \left\{ \left[
\begin{array}{cc}
 \xi I & O \\ 
 O & \mu I \\
\end{array}
\right] + \frac{\mu \nu - \xi \eta}{\gamma \eta - \beta \mu}
 \left[
\begin{array}{cc}
 -\gamma I & O \\ 
 O & \beta I \\
\end{array}
\right]
 \left[
\begin{array}{c}
 q_n \\
 -\mu I \\
\end{array}
\right]
(\mu \nu I - r_n q_n)^{-1}  
\left[
\begin{array}{cc}
 r_n & - \mu I \\
\end{array}
\right] \right\}
\nonumber \\[1mm]
& =  \mu \xi I, 
\label{sharp2}
\end{align}
we can rewrite (\ref{q_r_tilde1}) and (\ref{u_v_1}) as
\begin{align}
& 
 \left[
\begin{array}{c}
 \widetilde{q}_n \\
 -\xi I \\
\end{array}
\right]
(\xi \eta I - \widetilde{r}_n \widetilde{q}_n)^{-1}  
\left[
\begin{array}{cc}
 \widetilde{r}_n & - \xi I \\
\end{array}
\right]
\nonumber \\[2mm]
& = \frac{1}{\alpha \gamma}
 \left\{ \left[
\begin{array}{cc}
 \alpha I & O \\ 
 O & \gamma I \\
\end{array}
\right] + \frac{\alpha \beta - \gamma \delta}{\beta \xi - \gamma \nu}
 \left[
\begin{array}{c}
 u_{n+1} \\
 -\gamma I \\
\end{array}
\right]
(\gamma \delta I - v_{n+1} u_{n+1})^{-1}  
\left[
\begin{array}{cc}
 v_{n+1} & - \gamma I \\
\end{array}
\right] 
 \left[
\begin{array}{cc}
 \xi I & O \\ 
 O & -\nu I \\
\end{array}
\right]
\right\} 
\nonumber \\[1mm]
& \times 
 \left[
\begin{array}{c}
 q_n \\
 -\xi I \\
\end{array}
\right]
(\xi \eta I - r_n q_n)^{-1}  
\left[
\begin{array}{cc}
 r_n & - \xi I \\
\end{array}
\right]
\nonumber \\[1mm]
& \times 
\left\{ \left[
\begin{array}{cc}
 \gamma I & O \\ 
 O & \alpha I \\
\end{array}
\right] + \frac{\alpha \beta - \gamma \delta}{\alpha \nu - \delta \xi}
 \left[
\begin{array}{cc}
 \xi I & O \\ 
 O & -\nu I \\
\end{array}
\right]
 \left[
\begin{array}{c}
 u_n \\
 -\alpha I \\
\end{array}
\right]
(\alpha \beta I - v_n u_n)^{-1}  
\left[
\begin{array}{cc}
 v_n & - \alpha I \\
\end{array}
\right] \right\},
\label{q_r_tilde2} 
\end{align}
and 
\begin{align}
& \frac{1}{\mu \xi} 
 \left\{ \left[
\begin{array}{cc}
 \mu I & O \\ 
 O & \xi I \\
\end{array}
\right] + \frac{\mu \nu - \xi \eta}{\beta \xi - \gamma \nu}
 \left[
\begin{array}{c}
 \widetilde{q}_{n} \\
 -\xi I \\
\end{array}
\right]
(\xi \eta I - \widetilde{r}_{n} \widetilde{q}_{n})^{-1}  
\left[
\begin{array}{cc}
 \widetilde{r}_{n} & - \xi I \\
\end{array}
\right] 
 \left[
\begin{array}{cc}
 -\gamma I & O \\ 
 O & \beta I \\
\end{array}
\right]
\right\} 
\nonumber \\[1mm]
& \times 
 \left[
\begin{array}{c}
 u_n \\
 -\gamma I \\
\end{array}
\right]
(\gamma \delta I - v_n u_n)^{-1}  
\left[
\begin{array}{cc}
 v_n & - \gamma I \\
\end{array}
\right]
\nonumber \\[1mm]
& \times 
 \left\{ \left[
\begin{array}{cc}
 \xi I & O \\ 
 O & \mu I \\
\end{array}
\right] + \frac{\mu \nu - \xi \eta}{\gamma \eta - \beta \mu}
 \left[
\begin{array}{cc}
 -\gamma I & O \\ 
 O & \beta I \\
\end{array}
\right]
 \left[
\begin{array}{c}
 q_n \\
 -\mu I \\
\end{array}
\right]
(\mu \nu I - r_n q_n)^{-1}  
\left[
\begin{array}{cc}
 r_n & - \mu I \\
\end{array}
\right] \right\}
\nonumber \\[2mm]
& = 
 \left[
\begin{array}{c}
 u_{n+1} \\
 -\gamma I \\
\end{array}
\right]
(\gamma \delta I - v_{n+1} u_{n+1})^{-1}  
\left[
\begin{array}{cc}
 v_{n+1} & - \gamma I \\
\end{array}
\right],
\label{u_v_2} 
\end{align}
respectively. 
Thus, we can set 
\begin{align}
& 
 \left[
\begin{array}{c}
 \widetilde{q}_n \\
 -\xi I \\
\end{array}
\right]
\nonumber \\[2mm]
& = \left\{ \left[
\begin{array}{cc}
 \alpha I & O \\ 
 O & \gamma I \\
\end{array}
\right] + \frac{\alpha \beta - \gamma \delta}{\beta \xi - \gamma \nu}
 \left[
\begin{array}{c}
 u_{n+1} \\
 -\gamma I \\
\end{array}
\right]
(\gamma \delta I - v_{n+1} u_{n+1})^{-1}  
\left[
\begin{array}{cc}
 v_{n+1} & - \gamma I \\
\end{array}
\right] 
 \left[
\begin{array}{cc}
 \xi I & O \\ 
 O & -\nu I \\
\end{array}
\right]
\right\} 
\nonumber \\[1mm]
& \times 
 \left[
\begin{array}{c}
 q_n \\
 -\xi I \\
\end{array}
\right] X_n, 
\label{2.15} 
\end{align}
%
\begin{align}
& 
\left[
\begin{array}{cc}
 \widetilde{r}_n & - \xi I \\
\end{array}
\right]
= Y_n 
\left[
\begin{array}{cc}
 r_n & - \xi I \\
\end{array}
\right]
\nonumber \\[1mm]
& \times 
\left\{ \left[
\begin{array}{cc}
 \gamma I & O \\ 
 O & \alpha I \\
\end{array}
\right] + \frac{\alpha \beta - \gamma \delta}{\alpha \nu - \delta \xi}
 \left[
\begin{array}{cc}
 \xi I & O \\ 
 O & -\nu I \\
\end{array}
\right]
 \left[
\begin{array}{c}
 u_n \\
 -\alpha I \\
\end{array}
\right]
(\alpha \beta I - v_n u_n)^{-1}  
\left[
\begin{array}{cc}
 v_n & - \alpha I \\
\end{array}
\right] \right\},
\label{2.16} 
\end{align}
and 
\begin{align}
& \left[
\begin{array}{c}
 u_{n+1} \\
 -\gamma I \\
\end{array}
\right]
\nonumber \\[2mm]
& = \left\{ \left[
\begin{array}{cc}
 \mu I & O \\ 
 O & \xi I \\
\end{array}
\right] + \frac{\mu \nu - \xi \eta}{\beta \xi - \gamma \nu}
 \left[
\begin{array}{c}
 \widetilde{q}_{n} \\
 -\xi I \\
\end{array}
\right]
(\xi \eta I - \widetilde{r}_{n} \widetilde{q}_{n})^{-1}  
\left[
\begin{array}{cc}
 \widetilde{r}_{n} & - \xi I \\
\end{array}
\right] 
 \left[
\begin{array}{cc}
 -\gamma I & O \\ 
 O & \beta I \\
\end{array}
\right]
\right\} 
\nonumber \\[1mm]
& \times 
 \left[
\begin{array}{c}
 u_n \\
 -\gamma I \\
\end{array}
\right] Z_n, 
\label{2.17} 
\end{align}
\begin{align}
& \left[
\begin{array}{cc}
 v_{n+1} & - \gamma I \\
\end{array}
\right] = W_n 
\left[
\begin{array}{cc}
 v_n & - \gamma I \\
\end{array}
\right]
\nonumber \\[1mm]
& \times 
 \left\{ \left[
\begin{array}{cc}
 \xi I & O \\ 
 O & \mu I \\
\end{array}
\right] + \frac{\mu \nu - \xi \eta}{\gamma \eta - \beta \mu}
 \left[
\begin{array}{cc}
 -\gamma I & O \\ 
 O & \beta I \\
\end{array}
\right]
 \left[
\begin{array}{c}
 q_n \\
 -\mu I \\
\end{array}
\right]
(\mu \nu I - r_n q_n)^{-1}  
\left[
\begin{array}{cc}
 r_n & - \mu I \\
\end{array}
\right] \right\}, 
\label{2.18} 
\end{align}
where $X_n$, $Y_n$, $Z_n$ and $W_n$ are 
square matrices that satisfy the relations: 
\begin{equation}
X_n (\xi \eta I - \widetilde{r}_{n} \widetilde{q}_{n})^{-1} Y_n = \frac{1}{\alpha \gamma} (\xi \eta I - r_{n} q_{n})^{-1}, 
\label{2.19}
\end{equation}
and 
\begin{equation}
Z_n (\gamma \delta I - v_{n+1} u_{n+1})^{-1} W_n = \frac{1}{\mu \xi} (\gamma \delta I - v_{n} u_{n})^{-1}.
\label{2.20}
\end{equation}

Using (\ref{2.18}) and (\ref{2.20}), we can rewrite (\ref{2.15}) as
\begin{align}
& 
 \left[
\begin{array}{c}
 \widetilde{q}_n \\
 -\xi I \\
\end{array}
\right] X_n^{-1} 
= \left\{ \left[
\begin{array}{cc}
 \alpha I & O \\ 
 O & \gamma I \\
\end{array}
\right] + \frac{\alpha \beta - \gamma \delta}{\beta \xi - \gamma \nu}
 \left[
\begin{array}{c}
 u_{n+1} \\
 -\gamma I \\
\end{array}
\right]
\frac{1}{\mu \xi} Z_n^{-1}
(\gamma \delta I - v_{n} u_{n})^{-1}  
\left[
\begin{array}{cc}
 v_{n} & - \gamma I \\
\end{array}
\right] 
\right. 
\nonumber \\[1mm]
& \left. 
\times 
 \left[ \left[
\begin{array}{cc}
 \xi I & O \\ 
 O & \mu I \\
\end{array}
\right] + \frac{\mu \nu - \xi \eta}{\gamma \eta - \beta \mu}
 \left[
\begin{array}{cc}
 -\gamma I & O \\ 
 O & \beta I \\
\end{array}
\right]
 \left[
\begin{array}{c}
 q_n \\
 -\mu I \\
\end{array}
\right]
(\mu \nu I - r_n q_n)^{-1}  
\left[
\begin{array}{cc}
 r_n & - \mu I \\
\end{array}
\right] \right]
 \left[
\begin{array}{cc}
 \xi I & O \\ 
 O & -\nu I \\
\end{array}
\right]
\right\}
\nonumber \\[1mm]
& \times 
 \left[
\begin{array}{c}
 q_n \\
 -\xi I \\
\end{array}
\right] 
\nonumber \\[2mm]
& = 
\left[
\begin{array}{c}
 \alpha q_n \\
 - \gamma \xi I \\
\end{array}
\right] 
- \frac{\alpha \beta - \gamma \delta}{\beta \mu - \gamma \eta}
 \left[
\begin{array}{c}
 u_{n+1} \\
 -\gamma I \\
\end{array}
\right] Z_n^{-1}
(\gamma \delta I - v_n u_n)^{-1}  (\gamma \eta I - v_n q_n). 
\label{2.21} 
\end{align}
Similarly, using (\ref{2.16}) and (\ref{2.19}), we can rewrite (\ref{2.17}) as
\begin{align}
& \left[
\begin{array}{c}
 u_{n+1} \\
 -\gamma I \\
\end{array}
\right] Z_n^{-1} 
= \left\{ \left[
\begin{array}{cc}
 \mu I & O \\ 
 O & \xi I \\
\end{array}
\right] + \frac{\mu \nu - \xi \eta}{\beta \xi - \gamma \nu}
 \left[
\begin{array}{c}
 \widetilde{q}_{n} \\
 -\xi I \\
\end{array}
\right]
\frac{1}{\alpha \gamma} X_n^{-1} 
(\xi \eta I - r_{n} q_{n})^{-1}  
\left[
\begin{array}{cc}
 r_{n} & - \xi I \\
\end{array}
\right] \right.
\nonumber \\[1mm]
& \left. 
\times 
\left[ \left[
\begin{array}{cc}
 \gamma I & O \\ 
 O & \alpha I \\
\end{array}
\right] + \frac{\alpha \beta - \gamma \delta}{\alpha \nu - \delta \xi}
 \left[
\begin{array}{cc}
 \xi I & O \\ 
 O & -\nu I \\
\end{array}
\right]
 \left[
\begin{array}{c}
 u_n \\
 -\alpha I \\
\end{array}
\right]
(\alpha \beta I - v_n u_n)^{-1}  
\left[
\begin{array}{cc}
 v_n & - \alpha I \\
\end{array}
\right] \right]
 \left[
\begin{array}{cc}
 -\gamma I & O \\ 
 O & \beta I \\
\end{array}
\right]
\right\}
\nonumber \\[1mm]
& \times 
 \left[
\begin{array}{c}
 u_n \\
 -\gamma I \\
\end{array}
\right] 
\nonumber \\[2mm]
& = 
\left[
\begin{array}{c}
 \mu u_n \\
 -\gamma \xi I \\
\end{array}
\right] - \frac{\mu \nu - \xi \eta}{\alpha \nu - \delta \xi}
\left[
\begin{array}{c}
 \widetilde{q}_{n} \\
 -\xi I \\
\end{array}
\right] X_n^{-1} 
(\xi \eta I - r_{n} q_{n})^{-1} (\delta \xi I - r_{n} u_{n}). 
\label{2.22} 
\end{align}
Combining (\ref{2.21}) 
with (\ref{2.22}), we obtain 
\begin{align}
& 
 \left[
\begin{array}{c}
 \widetilde{q}_n \\
 -\xi I \\
\end{array}
\right] 
X_n^{-1} 
\left\{
I- \frac{(\alpha \beta - \gamma \delta)(\mu \nu - \xi \eta)}{(\beta \mu - \gamma \eta)(\alpha \nu - \delta \xi)}
(\xi \eta I-r_n q_n)^{-1} (\delta \xi I -r_n u_n) (\gamma \delta I - v_n u_n)^{-1} (\gamma \eta I -v_n q_n)
\right\}
\nonumber \\[2mm]
& = \left[
\begin{array}{c}
 \alpha q_n \\
 - \gamma \xi I \\
\end{array}
\right] 
- \frac{\alpha \beta - \gamma \delta}{\beta \mu - \gamma \eta}
 \left[
\begin{array}{c}
 \mu u_{n} \\
 -\gamma \xi I \\
\end{array}
\right] (\gamma \delta I - v_n u_n)^{-1}  (\gamma \eta I - v_n q_n), 
\label{2.23} 
\end{align}
and 
\begin{align}
& \left[
\begin{array}{c}
 u_{n+1} \\
 -\gamma I \\
\end{array}
\right] 
Z_n^{-1} 
\left\{
I- \frac{(\alpha \beta - \gamma \delta)(\mu \nu - \xi \eta)}{(\beta \mu - \gamma \eta)(\alpha \nu - \delta \xi)}
(\gamma \delta I - v_n u_n)^{-1}(\gamma \eta I -v_n q_n) (\xi \eta I-r_n q_n)^{-1} (\delta \xi I -r_n u_n)  
\right\}
\nonumber \\[2mm]
& = 
\left[
\begin{array}{c}
 \mu u_n \\
 -\gamma \xi I \\
\end{array}
\right] - \frac{\mu \nu - \xi \eta}{\alpha \nu - \delta \xi}
\left[
\begin{array}{c}
 \alpha q_{n} \\
 - \gamma \xi I \\
\end{array}
\right] (\xi \eta I - r_{n} q_{n})^{-1} (\delta \xi I - r_{n} u_{n}).
\label{2.24} 
\end{align}

Equations (\ref{2.23}) and (\ref{2.24}) imply 
\begin{align}
\widetilde{q}_n =& \; (\gamma \delta I - u_n v_n)^{-1}\left[
-\alpha \delta (\beta \mu - \gamma \eta) q_n + \mu \eta (\alpha \beta - \gamma \delta) u_n
- (\alpha \eta - \delta \mu) u_n v_n q_n \right]
\nonumber \\
& \times \left[ \beta \gamma (\alpha \eta - \delta \mu) I + (\beta \mu - \gamma \eta) v_n u_{n} 
 - (\alpha \beta - \gamma \delta) v_n q_n \right]^{-1} (\gamma \delta I - v_n u_n), 
\label{2.25} 
\end{align}
and 
\begin{align}
u_{n+1} =& \; (\xi \eta I - q_{n} r_{n})^{-1} \left[ 
 \mu \eta (\alpha \nu - \delta \xi) u_n - \alpha \delta (\mu \nu -\xi \eta) q_n 
- (\alpha \eta - \delta \mu) q_n r_n u_n \right]
\nonumber \\
& \times \left[ \nu \xi (\alpha \eta - \delta \mu) I - (\alpha \nu - \delta \xi)r_n q_n 
+ (\mu \nu - \xi \eta) r_{n} u_{n} \right]^{-1} (\xi \eta I - r_{n} q_{n}), 
\label{2.26} 
\end{align}
respectively, while equations (\ref{2.16}) and (\ref{2.18}) imply 
\begin{align}
\widetilde{r}_n =& \; (\alpha \beta I - v_n u_n) 
\left[ \alpha \delta (\beta \xi - \gamma \nu) I + (\alpha \nu - \delta \xi) v_n u_n 
 - (\alpha \beta - \gamma \delta) r_n u_n \right]^{-1} 
\nonumber \\
& \times \left[
-\beta \gamma (\alpha \nu - \delta \xi) r_n + \nu \xi (\alpha \beta - \gamma \delta) v_n
- (\beta \xi - \gamma \nu) r_n u_n v_n \right] (\alpha \beta I - u_n v_n)^{-1}, 
\label{2.29} 
\end{align}
and 
\begin{align}
v_{n+1} =& \; (\mu \nu I - r_{n} q_{n}) 
\left[ \mu \eta (\beta \xi - \gamma \nu) I - (\beta \mu - \gamma \eta)r_n q_n 
+ (\mu \nu - \xi \eta) v_{n} q_{n} \right]^{-1} 
\nonumber \\
& \times \left[ 
  \nu \xi (\beta \mu - \gamma \eta) v_n - \beta \gamma (\mu \nu -\xi \eta) r_n 
- (\beta \xi - \gamma \nu) v_n q_n r_n \right] (\mu \nu I - q_{n} r_{n})^{-1}, 
\label{2.30} 
\end{align}
respectively. 

With a lengthy (but 
almost straightforward) calculation, 
we can show 
that 
relation (\ref{2.19}) is a direct consequence of 
 (\ref{2.16}) and (\ref{2.23}), i.e., 
\begin{align}
& Y_n^{-1} (\widetilde{r}_{n} \widetilde{q}_{n} -\xi \eta I) X_n^{-1} (\xi \eta I - r_{n} q_{n})^{-1}
\nonumber \\[2mm]
& = Y_n^{-1}
\left[
\begin{array}{cc}
 \widetilde{r}_n & - \xi I \\
\end{array}
\right]
 \left[
\begin{array}{cc}
 I & O \\ 
 O & -\frac{\eta}{\xi} I \\
\end{array}
\right]
\left[
\begin{array}{c}
 \widetilde{q}_n \\
 -\xi I \\
\end{array}
\right] X_n^{-1} (\xi \eta I - r_{n} q_{n})^{-1}
\nonumber \\[2mm]
& = 
\left[
\begin{array}{cc}
 r_n & - \xi I \\
\end{array}
\right]
\left\{ \left[
\begin{array}{cc}
 \gamma I & O \\ 
 O & \alpha I \\
\end{array}
\right] + \frac{\alpha \beta - \gamma \delta}{\alpha \nu - \delta \xi}
 \left[
\begin{array}{cc}
 \xi I & O \\ 
 O & -\nu I \\
\end{array}
\right]
 \left[
\begin{array}{c}
 u_n \\
 -\alpha I \\
\end{array}
\right]
(\alpha \beta I - v_n u_n)^{-1}  
\left[
\begin{array}{cc}
 v_n & - \alpha I \\
\end{array}
\right] \right\}
\nonumber \\[1mm]
& \times
\left\{
\left[
\begin{array}{c}
 \alpha q_n \\
 \gamma \eta I \\
\end{array}
\right] 
- \frac{\alpha \beta - \gamma \delta}{\beta \mu - \gamma \eta}
 \left[
\begin{array}{c}
 \mu u_{n} \\
 \gamma \eta I \\
\end{array}
\right] (\gamma \delta I - v_n u_n)^{-1}  (\gamma \eta I - v_n q_n) \right\}
\nonumber \\[1mm]
& \times
\left\{
 \xi \eta I-r_n q_n - \frac{(\alpha \beta - \gamma \delta)(\mu \nu - \xi \eta)}{(\beta \mu - \gamma \eta)(\alpha \nu - \delta \xi)}
 (\delta \xi I -r_n u_n) (\gamma \delta I - v_n u_n)^{-1} (\gamma \eta I -v_n q_n)
\right\}^{-1}
\nonumber \\[2mm]
& = -\alpha \gamma I. 
\nonumber
\end{align}

Similarly, we can show that relation (\ref{2.20}) is a direct consequence of 
(\ref{2.18}) and (\ref{2.24}), i.e., 
\begin{align}
&
W_n^{-1} (v_{n+1} u_{n+1} - \gamma \delta I) Z_n^{-1} (\gamma \delta I - v_{n} u_{n})^{-1}
\nonumber \\[2mm]
&= W_n^{-1}
\left[
\begin{array}{cc}
 v_{n+1} & - \gamma I \\
\end{array}
\right]
\left[
\begin{array}{cc}
 I & O \\ 
 O & -\frac{\delta}{\gamma} I \\
\end{array}
\right]
\left[
\begin{array}{c}
 u_{n+1} \\
 -\gamma I \\
\end{array}
\right]  
Z_n^{-1} (\gamma \delta I - v_{n} u_{n})^{-1}
\nonumber \\[2mm]
&=
\left[
\begin{array}{cc}
 v_n & - \gamma I \\
\end{array}
\right]
 \left\{ \left[
\begin{array}{cc}
 \xi I & O \\ 
 O & \mu I \\
\end{array}
\right] + \frac{\mu \nu - \xi \eta}{\gamma \eta - \beta \mu}
 \left[
\begin{array}{cc}
 -\gamma I & O \\ 
 O & \beta I \\
\end{array}
\right]
 \left[
\begin{array}{c}
 q_n \\
 -\mu I \\
\end{array}
\right]
(\mu \nu I - r_n q_n)^{-1}  
\left[
\begin{array}{cc}
 r_n & - \mu I \\
\end{array}
\right] \right\}
\nonumber \\[1mm]
& \times \left\{
\left[
\begin{array}{c}
 \mu u_n \\
 \delta \xi I \\
\end{array}
\right] - \frac{\mu \nu - \xi \eta}{\alpha \nu - \delta \xi}
\left[
\begin{array}{c}
 \alpha q_{n} \\
 \delta \xi I \\
\end{array}
\right] (\xi \eta I - r_{n} q_{n})^{-1} (\delta \xi I - r_{n} u_{n})
\right\}
\nonumber \\[1mm]
& \times
\left\{ \gamma \delta I - v_n u_n
- \frac{(\alpha \beta - \gamma \delta)(\mu \nu - \xi \eta)}{(\beta \mu - \gamma \eta)(\alpha \nu - \delta \xi)}
(\gamma \eta I -v_n q_n) (\xi \eta I-r_n q_n)^{-1} (\delta \xi I -r_n u_n)  
\right\}^{-1} 
\nonumber \\[2mm]
& = -\mu \xi I.
\nonumber
\end{align}


To summarize, 
the discrete zero-curvature 
condition (\ref{fd-Lax}) for the Lax pair given by (\ref{Ln_YB}) and (\ref{Ln_YB2}) 
is equivalent to the fully discrete system of equations (\ref{2.25})--(\ref{2.30}). 


\subsection{Complex conjugation reduction and continuous limit}
\label{subsection2.4}

Note that (\ref{2.29}) and (\ref{2.30}) admit the alternative expressions: 
\begin{align}
\widetilde{r}_n =& \; (\gamma \delta I - v_n u_n)^{-1}\left[
- \beta \gamma (\alpha \nu - \delta \xi) r_n + \nu \xi (\alpha \beta - \gamma \delta) v_n
- (\beta \xi - \gamma \nu) v_n u_n r_n \right]
\nonumber \\
& \times \left[ \alpha \delta (\beta \xi - \gamma \nu) I + (\alpha \nu - \delta \xi) u_n v_n 
 - (\alpha \beta - \gamma \delta) u_n r_n \right]^{-1} (\gamma \delta I - u_n v_n), 
\label{2.27} 
\end{align}
and 
\begin{align}
v_{n+1} =& \; (\xi \eta I - r_{n} q_{n})^{-1} \left[ 
 \nu \xi  (\beta \mu - \gamma \eta) v_n - \beta \gamma (\mu \nu -\xi \eta) r_n 
- (\beta \xi - \gamma \nu) r_n q_n v_n \right]
\nonumber \\
& \times \left[ \mu \eta (\beta \xi - \gamma \nu) I - (\beta \mu - \gamma \eta)q_n r_n 
+ (\mu \nu - \xi \eta) q_{n} v_{n} \right]^{-1} (\xi \eta I - q_{n} r_{n}), 
\label{2.28} 
\end{align}
respectively. 
In fact, (\ref{2.27}) can be obtained from 
(\ref{2.16}) with (\ref{sharp1}), while (\ref{2.28}) can 
be 
obtained from 
(\ref{2.18}) with (\ref{sharp2}). 

In the case of \mbox{$\beta=\alpha^\ast$}, \mbox{$\delta=\gamma^\ast$}, 
\mbox{$\nu=\mu^\ast$} and \mbox{$\eta=\xi^\ast$}, 
the system of equations (\ref{2.25}), (\ref{2.26}), (\ref{2.27}) and (\ref{2.28}) admits the complex conjugation reduction; 
that is, the reduction 
\mbox{$r_n=\sigma q_n^\ast$}, \mbox{$v_n=\sigma u_n^\ast$} with a real constant $\sigma$
on the right-hand sides of the equations 
implies the reduction 
\mbox{$\widetilde{r}_n=\sigma \widetilde{q}_n^{\, \ast}$}, \mbox{$v_{n+1}=\sigma u_{n+1}^{\, \ast}$}
on the left-hand sides of the equations. 
However, this reduction is outside the scope of this paper. 



If the parameters satisfy the conditions: 
\begin{equation}
\gamma = \alpha^\ast, \hspace{5mm} \delta = \beta^\ast,  
\hspace{5mm} \xi = \mu^\ast, \hspace{5mm} \eta=\nu^\ast, 
\nonumber
\end{equation}
the fully discrete system of equations (\ref{2.25})--(\ref{2.30}) 
admits 
the Hermitian conjugation reduction; that is, 
the reduction 
\mbox{$r_n= q_n^\dagger$}, 
\mbox{$v_n=u_n^\dagger$} 
on the right-hand sides of the equations 
implies the reduction 
\mbox{$\widetilde{r}_n= \widetilde{q}_n^{\,
	\dagger}$}, 
\mbox{$v_{n+1}=u_{n+1}^{\, \dagger}$}
on the left-hand sides of the equations. 
In particular, 
if we set
\begin{equation}
\alpha=-\gamma=2\mathrm{i}/h, \hspace{5mm} \beta=\delta=1, 
\hspace{5mm} \mu=\xi=1, \hspace{5mm} \nu=-\eta=2\mathrm{i}/\varDelta, 
\hspace{5mm} h, \varDelta \in \mathbb{R}_{\ne 0},  
\label{concrete_value}
\end{equation}
and impose 
this reduction, 
we obtain 
\begin{align}
\widetilde{q}_n =& \; \left( I -\frac{\mathrm{i}h}{2}u_n u_n^\dagger \right)^{-1}\left[
 \left( 1+ \frac{h\varDelta}{4} \right) q_n + \mathrm{i} h u_n 
 - \frac{\mathrm{i}h}{2} \left( 1 - \frac{h \varDelta}{4} \right) u_n u_n^\dagger q_n \right]
\nonumber \\
& \times \left[ \left( 1- \frac{h\varDelta}{4} \right) I 
 + \frac{\mathrm{i}h}{2} \left( 1 + \frac{h\varDelta}{4} \right) u_n^\dagger u_{n} 
 +\frac{h\varDelta}{2} u_n^\dagger q_n \right]^{-1} \left( I -\frac{\mathrm{i}h}{2} u_n^\dagger u_n\right), 
\label{} 
\end{align}
\begin{align}
u_{n+1} =& \; \left( I - \frac{\mathrm{i}\varDelta}{2}q_{n} q_{n}^\dagger \right)^{-1} \left[ 
 \left( 1+\frac{h\varDelta}{4}\right) u_n - \mathrm{i}\varDelta q_n 
 +\frac{\mathrm{i}\varDelta}{2} \left( 1 - \frac{h\varDelta}{4} \right) q_n q_n^\dagger u_n \right]
\nonumber \\
& \times \left[ \left( 1-\frac{h\varDelta}{4} \right) I 
- \frac{\mathrm{i}\varDelta}{2} \left( 1+\frac{h\varDelta}{4}\right)q_n^\dagger q_n 
+ \frac{h\varDelta}{2} q_{n}^\dagger u_{n} \right]^{-1} \left( I - \frac{\mathrm{i}\varDelta}{2} q_{n}^\dagger q_{n} \right), 
\label{} 
\end{align}
which 
can be simplified to 
\begin{equation} 
\label{fdMTM1}
\left\{ 
\begin{split}
& \widetilde{q}_n = \frac{ \left( 1+ \frac{h\varDelta}{4} \right) q_n + \mathrm{i} h u_n 
 - \frac{\mathrm{i}h}{2} \left( 1 - \frac{h \varDelta}{4} \right) |u_n|^2 q_n}
{\left( 1- \frac{h\varDelta}{4} \right)
 + \frac{\mathrm{i}h}{2} \left( 1 + \frac{h\varDelta}{4} \right) |u_n|^2 
 +\frac{h\varDelta}{2} u_n^\ast q_n}, 
\\[1.5mm]
& u_{n+1} = \frac{\left( 1+\frac{h\varDelta}{4}\right) u_n - \mathrm{i}\varDelta q_n 
 +\frac{\mathrm{i}\varDelta}{2} \left( 1 - \frac{h\varDelta}{4} \right) |q_n|^2 u_n}
{\left( 1-\frac{h\varDelta}{4} \right)
- \frac{\mathrm{i}\varDelta}{2} \left( 1+\frac{h\varDelta}{4}\right) |q_n|^2 
+ \frac{h\varDelta}{2} q_{n}^\ast u_{n}},
\end{split} 
\right. 
\end{equation}
in the case of the scalar dependent variables. Note that (\ref{fdMTM1}) can be rewritten as 
\begin{equation} 
\label{fdMTM2}
\left\{ 
\begin{split}
& \mathrm{i} \frac{\widetilde{q}_n -q_n}{h} = \frac{\frac{\mathrm{i}\varDelta}{2} q_n -u_n 
 +|u_n|^2 q_n -\frac{\mathrm{i}\varDelta}{2} u_n^\ast q_n^2}
{\left( 1- \frac{h\varDelta}{4} \right)
 + \frac{\mathrm{i}h}{2} \left( 1 + \frac{h\varDelta}{4} \right) |u_n|^2 
 +\frac{h\varDelta}{2} u_n^\ast q_n}, 
\\[1.5mm]
& \mathrm{i} \frac{u_{n+1}-u_n}{\varDelta} = \frac{\frac{\mathrm{i}h}{2} u_n +q_n 
 -|q_n|^2 u_n -\frac{\mathrm{i}h}{2}q_{n}^\ast u_{n}^2}
{\left( 1-\frac{h\varDelta}{4} \right)
- \frac{\mathrm{i}\varDelta}{2} \left( 1+\frac{h\varDelta}{4}\right) |q_n|^2 
+ \frac{h\varDelta}{2} q_{n}^\ast u_{n}},
\end{split} 
\right. 
\end{equation}
which can be interpreted as 
a fully discrete analog of the massive Thirring model in light-cone coordinates (\ref{cMTM}). 
We mention 
that 
another 
discrete analog 
of  
the massive Thirring model in light-cone coordinates 
was proposed 
in~\cite{NCQ83}. 

In the continuous limit of time (\mbox{$h \to 0$}), 
the fully discrete massive Thirring model (\ref{fdMTM2}) reduces to 
the semi-discrete massive Thirring model: 
\begin{equation} 
\label{semiMTM1}
\left\{ 
\begin{split}
& \mathrm{i} q_{n,t} = \frac{\mathrm{i}\varDelta}{2} q_n -u_n 
 +|u_n|^2 q_n -\frac{\mathrm{i}\varDelta}{2} u_n^\ast q_n^2,
\\[1.5mm]
& \mathrm{i} \frac{u_{n+1}-u_n}{\varDelta} = \frac{q_n -|q_n|^2 u_n}
{1- \frac{\mathrm{i}\varDelta}{2} |q_n|^2}, 
\end{split} 
\right. 
\end{equation}
which is an integrable 
space-discretization 
of the massive Thirring model in light-cone coordinates (\ref{cMTM}); 
with 
minor manipulation, (\ref{semiMTM1}) can be identified 
with the semi-discrete massive Thirring model 
proposed in our previous paper 
(see (3.3) in~\cite{Me2015}).

\section{B\"acklund--Darboux transformation}

We present 
the binary B\"acklund--Darboux transformation 
for the 
discrete massive Thirring model 
given by equations 
(\ref{2.25})--(\ref{2.30}); 
note that 
Xu and Pelinovsky~\cite{Pelinovsky2019}
obtained some 
relevant results 
previously.

Because the Lax matrices 
(\ref{Ln_YB}) and (\ref{Ln_YB2}) 
have their origin in 
the binary B\"acklund--Darboux transformation 
for the continuous 
Chen--Lee--Liu hierarchy, 
the binary B\"acklund--Darboux transformation 
for the 
discrete massive Thirring model 
naturally has 
the same form. 
We use the hat  $\;\widehat{}\;$  
to denote the B\"acklund--Darboux transformation:  
\begin{equation}
\left[
\begin{array}{c}
 \widehat{\Psi}_{1, n}  \\
 \widehat{\Psi}_{2, n} \\
\end{array}
\right]
= {\cal L}({\cal U}_n, {\cal V}_n; a, b, c, d)
\left[
\begin{array}{c}
 \Psi_{1,n}  \\
 \Psi_{2,n} \\
\end{array}
\right],
\label{DBtrans}
\end{equation}
where the 
Darboux
matrix
${\cal L}({\cal U}_n, {\cal V}_n; a, b, c, d)$  
is defined 
exactly in the same manner 
as 
in 
(\ref{Ln_YB}) and (\ref{Ln_YB2}): 
\begin{align}
& 
{\cal L} ({\cal U}_n, {\cal V}_n; a, b, c, d) 
\nonumber \\[2mm]
&:=  \left[
\begin{array}{cc}
 a I & O \\ 
 O & c I \\
\end{array}
\right] + \frac{ab-cd}{c \zeta^2 + b}
 \left[
\begin{array}{cc}
 {\cal U}_n {\cal V}_n (cd I - {\cal U}_n {\cal V}_n )^{-1} &  -c \zeta {\cal U}_n (cd I - {\cal V}_n {\cal U}_n)^{-1}  \\
 -c \zeta {\cal V}_n (cd I - {\cal U}_n {\cal V}_n )^{-1} &  c^2 \zeta^2 (cd I - {\cal V}_n {\cal U}_n)^{-1}  \\
\end{array}
\right]
\nonumber \\[2mm]
&\hphantom{:}=  \left[
\begin{array}{cc}
 a I & O \\ 
 O & c I \\
\end{array}
\right] + \frac{ab-cd}{c \zeta^2 + b}
 \left[
\begin{array}{c}
 {\cal U}_n \\
 -c \zeta I \\
\end{array}
\right]
(cd I - {\cal V}_n {\cal U}_n)^{-1}  
\left[
\begin{array}{cc}
 {\cal V}_n & - c \zeta I \\
\end{array}
\right]. 
\label{Ln_DarBa1}
\end{align}
Here, $a$, $b$, $c$ and $d$
are nonzero 
constant parameters 
satisfying the condition 
\mbox{$ab-cd \neq 0$}.  
Similarly to (\ref{Ln_YB3}), 
${\cal L} ({\cal U}_n, {\cal V}_n; a, b, c, d)$ can be rewritten 
up to an overall factor 
as
\begin{align}
& 
\frac{\zeta+\frac{b}{c\zeta}}{\frac{a}{d}\zeta+\frac{1}{\zeta}}
{\cal L} ({\cal U}_n, {\cal V}_n; a, b, c, d) 
\nonumber \\[2mm]
&\hphantom{:}=  \left[
\begin{array}{cc}
 d I & O \\ 
 O & b I \\
\end{array}
\right] + \frac{ab-cd}{a + \frac{d}{\zeta^2}}
 \left[
\begin{array}{c}
 -\frac{d}{\zeta} I \\
 {\cal V}_n \\
\end{array}
\right]
(cd I - {\cal U}_n {\cal V}_n)^{-1}  
\left[
\begin{array}{cc}
 -\frac{d}{\zeta} I & {\cal U}_n \\
\end{array}
\right]. 
\label{Ln_DarBa2}
\end{align}
The definition (\ref{Ln_DarBa1}) and 
the alternative expression (\ref{Ln_DarBa2}) together with the linear 
problem (\ref{DBtrans}) 
imply that the unknown functions ${\cal U}_n$ and ${\cal V}_n$ 
can be expressed 
in terms of 
the 
linear 
eigenfunction 
as 
\begin{equation}
 {\cal U}_n= -a \left. (\zeta \Psi_{1,n}^{(1)}\Psi_{2,n}^{(1)\,-1}) \right|_{\zeta^2=-\frac{d}{a}}, \hspace{5mm} 
 {\cal V}_n= c \left. (\zeta \Psi_{2,n}^{(2)}\Psi_{1,n}^{(2)\,-1}) \right|_{\zeta^2=-\frac{b}{c}}, 
\label{intermediate_def}
\end{equation}
respectively. 
The superscript ${}^{(j)}\; (j=1,2)$ of the linear eigenfunction 
is used to stress that we can choose 
different 
linear 
eigenfunctions to define ${\cal U}_n$ and ${\cal V}_n$. 
Note that 
with an appropriate choice of the boundary/initial conditions, 
the 
functions $\zeta \Psi_{1,n}^{(1)}\Psi_{2,n}^{(1)\,-1}$ and $\zeta \Psi_{2,n}^{(2)}\Psi_{1,n}^{(2)\,-1}$ 
can be chosen as even functions in $\zeta$ (see (\ref{gDB2}) with (\ref{Ln_YB}) or
(\ref{gDB_time}) with (\ref{Ln_YB2})). 

The compatibility condition for 
the overdetermined 
linear 
equations (\ref{gDB2}) and (\ref{DBtrans}) 
is given by
%
\begin{equation}
 {\cal L}  (\widehat{q}_n, \widehat{r}_n; \mu, \nu, \xi, \eta) {\cal L} ({\cal U}_n, {\cal V}_n; a, b, c, d) 
 = {\cal L} ({\cal U}_{n+1}, {\cal V}_{n+1}; a, b, c, d)  {\cal L}  (q_n, r_n; \mu, \nu, \xi, \eta), 
\label{L_D_com}
\end{equation}
while the compatibility condition for 
the overdetermined 
linear 
equations (\ref{gDB_time}) and (\ref{DBtrans}) is given by
\begin{equation}
 {\cal L} (\widetilde{{\cal U}}_n, \widetilde{{\cal V}}_n; a, b, c, d) {\cal L}(u_n, v_n; \alpha, \beta, \gamma, \delta)
 =  {\cal L}(\widehat{u}_{n}, \widehat{v}_{n}; \alpha, \beta, \gamma, \delta) {\cal L} ({\cal U}_n, {\cal V}_n; a, b, c, d).
\label{V_D_com}
\end{equation}
Here, we assume 
that 
$\nu/\xi$, $\eta/\mu$, $\beta/\gamma$, $\delta/\alpha$, $b/c$ and $d/a$ 
are 
pairwise distinct. 

Because equations (\ref{L_D_com}) and (\ref{V_D_com}) 
have 
the same form as 
the discrete zero-curvature condition 
(\ref{fd-Lax}), 
we can solve these equations 
in the same way as 
in subsection~\ref{subsec2.3}. 
Thus, 
we can express $\widehat{q}_n$, $\widehat{r}_n$, $\widehat{u}_{n}$ and $\widehat{v}_{n}$ 
in terms of $q_n$, $r_n$, $u_n$, $v_n$, ${\cal U}_n$ and ${\cal V}_n$. 
The expressions 
are given 
by 
equations (\ref{2.25})--(\ref{2.30}) 
with an appropriate replacement 
of the parameters, variables and notation, i.e., 
\begin{align}
\widehat{q}_n =& \; (cd I - {\cal U}_n {\cal V}_n)^{-1}\left[
-ad (b \mu - c \eta) q_n + \mu \eta (ab - cd) {\cal U}_n
- (a \eta - d \mu) {\cal U}_n {\cal V}_n q_n \right]
\nonumber \\
& \times \left[ b c (a \eta - d \mu) I + (b \mu - c \eta) {\cal V}_n {\cal U}_{n} 
 - (ab - cd) {\cal V}_n q_n \right]^{-1} (cd I - {\cal V}_n {\cal U}_n), 
\label{BD_q} 
\\[1mm]
\widehat{r}_n =& \; (ab I - {\cal V}_n {\cal U}_n) 
\left[ ad (b \xi - c \nu) I + (a \nu - d \xi) {\cal V}_n {\cal U}_n 
 - (ab - cd) r_n {\cal U}_n \right]^{-1} 
\nonumber \\
& \times \left[
-bc (a \nu - d \xi) r_n + \nu \xi (ab - cd) {\cal V}_n
- (b \xi - c \nu) r_n {\cal U}_n {\cal V}_n \right] (ab I - {\cal U}_n {\cal V}_n)^{-1}, 
\label{BD_r} 
\\[1mm]
\widehat{u}_{n} =& \; (cd I - {\cal U}_{n} {\cal V}_{n})^{-1} \left[
  ad (\alpha b - \delta c) u_n - \alpha \delta (ab -cd) {\cal U}_n 
- (\alpha d - \delta a) {\cal U}_n {\cal V}_n u_n \right]
\nonumber \\
& \times \left[ bc (\alpha d - \delta a) I - (\alpha b - \delta c){\cal V}_n {\cal U}_n 
+ (ab - cd) {\cal V}_{n} u_{n} \right]^{-1} (cd I - {\cal V}_{n} {\cal U}_{n}), 
\label{BD_u} 
\\[1mm]
\widehat{v}_{n} =& \; (ab I - {\cal V}_{n} {\cal U}_{n}) 
\left[ ad (\beta c - \gamma b) I - (\beta a - \gamma d){\cal V}_n {\cal U}_n 
+ (ab - cd) v_{n} {\cal U}_{n} \right]^{-1} 
\nonumber \\
& \times \left[
  bc (\beta a - \gamma d) v_n - \beta \gamma (ab -cd) {\cal V}_n 
- (\beta c - \gamma b) v_n {\cal U}_n {\cal V}_n \right] (ab I - {\cal U}_{n} {\cal V}_{n})^{-1}. 
\label{BD_v} 
\end{align}

As discussed in subsection~\ref{subsection2.4}, 
if the parameters satisfy the conditions: 
\begin{equation}
\gamma = \alpha^\ast, \hspace{5mm} \delta = \beta^\ast,  
\hspace{5mm} \xi = \mu^\ast, \hspace{5mm} \eta=\nu^\ast, 
\nonumber
\end{equation}
the fully discrete system of equations (\ref{2.25})--(\ref{2.30}) 
admits 
the Hermitian conjugation reduction; the B\"acklund--Darboux transformation 
formulas (\ref{BD_q})--(\ref{BD_v}) are compatible with this reduction 
if we further impose 
the conditions:
\begin{equation}
c = a^\ast, \hspace{5mm} d = b^\ast, \hspace{5mm} {\cal V}_{n}={\cal U}_{n}^\dagger. 
\nonumber
\end{equation}
That is, 
the reduction 
\mbox{$r_n= q_n^\dagger$}, 
\mbox{$v_n=u_n^\dagger$}
on the right-hand sides of the equations 
implies the reduction 
\mbox{$\widehat{r}_n= \widehat{q}_n^{\,
	\dagger}$}, 
\mbox{$\widehat{v}_{n}=\widehat{u}_{n}^{\, \dagger}$}
on the left-hand sides of the equations. 

In the remaining part of this section, we
consider the 
case of the scalar dependent variables 
and choose the values of the parameters as given in 
(\ref{concrete_value}). 
Thus, the Hermitian conjugation reduction is simplified to 
the complex conjugation reduction; 
the Lax-pair representation 
and 
the formulas 
for the B\"acklund--Darboux transformation 
can be summarized 
as follows. 

%
The compatibility condition for
the 
overdetermined 
linear equations: 
\begin{equation}
\left[
\begin{array}{c}
 \Psi_{1, n+1}  \\
 \Psi_{2, n+1} \\
\end{array}
\right]
= \left\{
I + \frac{\mathrm{i}\varDelta}{\left( 1-\frac{\mathrm{i}\varDelta}{2}\zeta^2 \right) 
	\left( 1 - \frac{\mathrm{i}\varDelta}{2}  |q_n|^2 \right)}
 \left[
\begin{array}{cc}
 |q_n|^2 &  -\zeta q_n \\
 - \zeta q_n^\ast & \zeta^2 \\
\end{array}
\right]
\right\}
\left[
\begin{array}{c}
 \Psi_{1,n}  \\
 \Psi_{2,n} \\
\end{array}
\right], 
\label{sMTM_L}
\end{equation}
and 
\begin{equation}
\left[
\begin{array}{c}
 \widetilde{\Psi}_{1, n}  \\
 \widetilde{\Psi}_{2, n} \\
\end{array}
\right]
= \left\{ 
 I + \frac{\mathrm{i} h}{\left( 1 - \frac{\mathrm{i}h}{2} \frac{1}{\zeta^2}\right)
	\left( 1 - \frac{\mathrm{i}h}{2} |u_n|^2 \right)}
 \left[
\begin{array}{cc}
 \frac{1}{\zeta^2} &  -\frac{1}{\zeta} u_n  \\[1mm]
 -\frac{1}{\zeta} u_n^\ast  &  |u_n|^2  \\
\end{array}
\right]
\right\}
\left[
\begin{array}{c}
 \Psi_{1,n}  \\
 \Psi_{2,n} \\
\end{array}
\right],
\label{sMTM_V}
\end{equation}
is equivalent to 
the fully discrete 
massive Thirring model 
in light-cone coordinates 
(\ref{fdMTM1}) (or (\ref{fdMTM2})). 
Here, we 
rescaled the temporal Lax matrix 
as in (\ref{Ln_YB3}). 
This 
Lax-pair representation 
is form-invariant 
under the binary B\"acklund--Darboux transformation: 
\begin{equation}
\left[
\begin{array}{c}
 \widehat{\Psi}_{1, n}  \\
 \widehat{\Psi}_{2, n} \\
\end{array}
\right]
= \left\{
\left[
\begin{array}{cc}
 a & 0 \\ 
 0 & a^\ast\\
\end{array}
\right] + \frac{ab-a^\ast b^\ast}{(a^\ast \zeta^2 + b) \left( a^\ast b^\ast - |{\cal U}_n|^2 \right)}
 \left[
\begin{array}{cc}
 |{\cal U}_n|^2 &  -a^\ast \zeta {\cal U}_n \\
 -a^\ast \zeta {\cal U}_n^{\ast} &  a^{*\hspace{1pt} 2} \zeta^2 \\
\end{array}
\right]
\right\}
\left[
\begin{array}{c}
 \Psi_{1,n}  \\
 \Psi_{2,n} \\
\end{array}
\right],
\nonumber
\end{equation}
where the unknown function ${\cal U}_n$ is 
expressed in terms of the linear eigenfunction 
as 
\begin{equation}
 {\cal U}_n= -a \left. (\zeta \Psi_{1,n} \Psi_{2,n}^{-1}) \right|_{\zeta^2=-\frac{b^\ast}{a}}, 
\label{U_exp}
\end{equation}
and the transformed variables are given by 
\begin{align}
\widehat{q}_n =& \; \frac{
-ab^\ast ( 2\mathrm{i}a^\ast \! + \varDelta b ) q_n - 2\mathrm{i} (ab - a^\ast b^\ast) {\cal U}_n
+ (2\mathrm{i} a + \varDelta b^\ast) |{\cal U}_n|^2 q_n}{-a^\ast b (2\mathrm{i} a + \varDelta b^\ast) 
 + (2\mathrm{i} a^\ast \! +\varDelta b ) |{\cal U}_{n}|^2 - \varDelta (ab - a^\ast b^\ast) {\cal U}_n^{\ast} q_n}, 
\label{sBD_q} 
\\[1mm]
\widehat{u}_{n} =& \; \frac{ 
  ab^\ast (2\mathrm{i} b - h a^\ast) u_n - 2\mathrm{i} (ab -a^\ast b^\ast) {\cal U}_n 
- (2\mathrm{i} b^\ast \! - ha) |{\cal U}_n|^2 u_n}{a^\ast b(2\mathrm{i} b^\ast \! - h a) 
- (2\mathrm{i} b - ha^\ast) |{\cal U}_n|^2 
+ h (ab - a^\ast b^\ast) {\cal U}_{n}^{\ast} u_{n}}.
\label{sBD_u} 
\end{align}

We start with the trivial zero solution \mbox{$q_n=u_n=0$} for all 
time 
\mbox{$m \in {\mathbb Z}$} 
and 
choose the solution of the linear equations 
(\ref{sMTM_L}) and (\ref{sMTM_V}) as 
\begin{equation}
\left[
\begin{array}{c}
 \Psi_{1,n}(m) \\
 \Psi_{2,n}(m)\\
\end{array}
\right] 
= \left[
\begin{array}{c}
 k \zeta \left( \frac{1+\frac{\mathrm{i}h}{2}\frac{1}{\zeta^2}}{1-\frac{\mathrm{i}h}{2}\frac{1}{\zeta^2}}\right)^m \\[3mm]
 \left( \frac{1+\frac{\mathrm{i}\varDelta}{2}\zeta^2}{1-\frac{\mathrm{i}\varDelta}{2}\zeta^2}\right)^n \\
\end{array}
\right],  
\nonumber
\end{equation}
where $k$ 
is a 
nonzero complex 
constant. 
For convenience, we set \mbox{$k=\kappa a^\ast /b^\ast$}, 
where $\kappa$
is a 
nonzero 
complex 
constant. 
Then, 
(\ref{U_exp}) implies that 
\begin{equation}
 {\cal U}_n= \kappa a^\ast 
	\left( \frac{1+\frac{\mathrm{i}\varDelta b^\ast}{2a}}{1-\frac{\mathrm{i}\varDelta b^\ast}{2a}}\right)^n
	\left( \frac{1-\frac{\mathrm{i}ha}{2b^\ast}}{1+\frac{\mathrm{i}ha}{2b^\ast}}\right)^m, 
\nonumber
\end{equation}
so we obtain from (\ref{sBD_q}) and (\ref{sBD_u}) 
the one-soliton solution of 
(\ref{fdMTM1}) (or (\ref{fdMTM2})): 
\begin{align}
\widehat{q}_n =& \; \frac{
 \kappa \left( 1 - \frac{a^\ast b^\ast }{ab}\right) 
	\left( \frac{1+\frac{\mathrm{i}\varDelta b^\ast}{2a}}{1-\frac{\mathrm{i}\varDelta b^\ast}{2a}}\right)^n
	\left( \frac{1-\frac{\mathrm{i}ha}{2b^\ast}}{1+\frac{\mathrm{i}ha}{2b^\ast}}\right)^m
}{\left( 1 - \frac{\mathrm{i} \varDelta b^\ast}{2a} \right) 
 - \frac{a^\ast}{b} \left(1 - \frac{\mathrm{i}\varDelta b}{2a^\ast}\right) 
	|\kappa|^2 \left| \frac{1+\frac{\mathrm{i}\varDelta b^\ast}{2a}}{1-\frac{\mathrm{i}\varDelta b^\ast}{2a}}\right|^{2n}
	\left| \frac{1-\frac{\mathrm{i}ha}{2b^\ast}}{1+\frac{\mathrm{i}ha}{2b^\ast}}\right|^{2m}}, 
\nonumber
\\[1mm]
\widehat{u}_{n} =& \; \frac{ \kappa
 \left( 1 -\frac{a^\ast b^\ast }{ab}\right) 
	\left( \frac{1+\frac{\mathrm{i}\varDelta b^\ast}{2a}}{1-\frac{\mathrm{i}\varDelta b^\ast}{2a}}\right)^n
	\left( \frac{1-\frac{\mathrm{i}ha}{2b^\ast}}{1+\frac{\mathrm{i}ha}{2b^\ast}}\right)^m}
{ -\frac{b^\ast}{a} \left( 1+ \frac{\mathrm{i}ha}{2b^\ast} \right) 
 + \left( 1 + \frac{\mathrm{i}h a^\ast}{2b}\right) |\kappa|^2 \left| \frac{1+\frac{\mathrm{i}\varDelta b^\ast}{2a}}{1-\frac{\mathrm{i}\varDelta b^\ast}{2a}}\right|^{2n}
	\left| \frac{1-\frac{\mathrm{i}ha}{2b^\ast}}{1+\frac{\mathrm{i}ha}{2b^\ast}}\right|^{2m}}.
\nonumber
\end{align}




\section{Yang--Baxter map}
The 
discrete Lax matrix 
${\cal L} (q_n, r_n; \mu, \nu, \xi, \eta)$ 
defined in (\ref{Ln_YB}) 
has 
the property: 
\begin{equation}
{\cal L} (q_n, r_n; \mu, \nu, \xi, \eta) {\cal L} (q_n, r_n; \xi, \eta, \mu, \nu) = \mu \xi I.
\label{L_qr_inverse}
\end{equation}
Similarly, the discrete Lax matrix
${\cal L} (u_n, v_n; \alpha, \beta, \gamma, \delta)$ defined in (\ref{Ln_YB2}) 
has the property: 
\begin{equation}
{\cal L} (u_n, v_n; \alpha, \beta, \gamma, \delta) {\cal L} (u_n, v_n; \gamma, \delta, \alpha, \beta)  = \alpha \gamma I.
\label{L_uv_inverse}
\end{equation}

Using the 
property (\ref{L_qr_inverse}), 
we can rewrite the 
discrete zero-curvature 
condition (\ref{fd-Lax}) as 
\begin{equation}
{\cal L}  (\widetilde{q}_n, \widetilde{r}_n; \xi, \eta, \mu, \nu)  {\cal L}(u_{n+1}, v_{n+1}; \alpha, \beta, \gamma, \delta) = 
  {\cal L}(u_n, v_n; \alpha, \beta, \gamma, \delta)
  {\cal L}  (q_n, r_n; \xi, \eta, \mu, \nu).
\nonumber
\end{equation}
Thus, the  
map 
\mbox{$R_{\{\mu, \nu, \xi, \eta\}}^{\{\alpha, \beta, \gamma, \delta\}}:\ 
\left( q_n, r_n, u_{n+1}, v_{n+1} \right) \mapsto \left( \widetilde{q}_n, \widetilde{r}_n, u_n, v_n \right)$} 
takes exactly the same form as the map 
\mbox{$\left( q_n, r_n, u_n, v_n \right) \mapsto \left( \widetilde{q}_n, \widetilde{r}_n, u_{n+1}, v_{n+1} \right)$} 
described by (\ref{2.25})--(\ref{2.30}), 
but 
with
the interchange of 
the parameters: \mbox{$\mu \leftrightarrow \xi$} and \mbox{$\nu \leftrightarrow \eta$}. 
Hence, the 
map 
\mbox{$R_{\{\mu, \nu, \xi, \eta\}}^{\{\alpha, \beta, \gamma, \delta\}}$} 
can be written explicitly as 
\begin{align}
\widetilde{q}_n =& \; (\gamma \delta I - u_{n+1} v_{n+1})^{-1}\left[
-\alpha \delta (\beta \xi - \gamma \nu) q_n + \nu \xi (\alpha \beta - \gamma \delta) u_{n+1}
- (\alpha \nu - \delta \xi) u_{n+1} v_{n+1} q_n \right]
\nonumber \\
& \times \left[ \beta \gamma (\alpha \nu - \delta \xi) I + (\beta \xi - \gamma \nu) v_{n+1} u_{n+1} 
 - (\alpha \beta - \gamma \delta) v_{n+1} q_{n} \right]^{-1} (\gamma \delta I - v_{n+1} u_{n+1}), 
\label{YB_q1} 
\end{align}
\begin{align}
\widetilde{r}_n =& \; (\alpha \beta I - v_{n+1} u_{n+1}) 
\left[ \alpha \delta (\beta \mu - \gamma \eta) I + (\alpha \eta - \delta \mu) v_{n+1} u_{n+1} 
 - (\alpha \beta - \gamma \delta) r_n u_{n+1} \right]^{-1} 
\nonumber \\
& \times \left[ 
-\beta \gamma (\alpha \eta - \delta \mu) r_n + \mu \eta (\alpha \beta - \gamma \delta) v_{n+1}
- (\beta \mu - \gamma \eta) r_n u_{n+1} v_{n+1} \right] (\alpha \beta I - u_{n+1} v_{n+1})^{-1}, 
\label{YB_r1} 
\end{align}
\begin{align}
u_{n} =& \; (\mu \nu I - q_{n} r_{n})^{-1} \left[
 \nu \xi (\alpha \eta - \delta \mu) u_{n+1} + \alpha \delta (\mu \nu -\xi \eta) q_n 
- (\alpha \nu - \delta \xi) q_n r_n u_{n+1} \right]
\nonumber \\
& \times \left[ \mu \eta (\alpha \nu - \delta \xi) I - (\alpha \eta - \delta \mu)r_n q_n 
- (\mu \nu - \xi \eta) r_{n} u_{n+1} \right]^{-1} (\mu \nu I - r_{n} q_{n}), 
\label{YB_u1} 
\end{align}
\begin{align}
v_{n} =& \; (\xi \eta I - r_{n} q_{n}) 
\left[ \nu \xi (\beta \mu - \gamma \eta) I - (\beta \xi - \gamma \nu)r_n q_n 
- (\mu \nu - \xi \eta) v_{n+1} q_{n} \right]^{-1} 
\nonumber \\
& \times \left[
 \mu \eta (\beta \xi - \gamma \nu) v_{n+1} + \beta \gamma (\mu \nu -\xi \eta) r_n 
- (\beta \mu - \gamma \eta) v_{n+1} q_n r_n \right] (\xi \eta I - q_{n} r_{n})^{-1}. 
\label{YB_v1} 
\end{align}
As 
stated 
in subsection~\ref{subsec2.2}, 
the 
nonzero 
constant parameters 
$\mu$, $\nu$, $\xi$, $\eta$, $\alpha$, $\beta$, $\gamma$ and $\delta$
are 
required 
to 
satisfy 
the conditions \mbox{$\mu \nu - \xi \eta \neq 0$}, 
\mbox{$\alpha \beta - \gamma \delta \neq 0$}, 
%
\mbox{$\alpha \nu - \delta \xi \neq 0$}, \mbox{$\alpha \eta - \delta \mu \neq 0$}, 
\mbox{$\beta \mu - \gamma \eta \neq 0$} and \mbox{$\beta \xi - \gamma \nu \neq 0$}.
In other words, 
the values of $\nu/\xi$, $\eta/\mu$, $\beta/\gamma$ and $\delta/\alpha$ 
are 
pairwise distinct. 
%
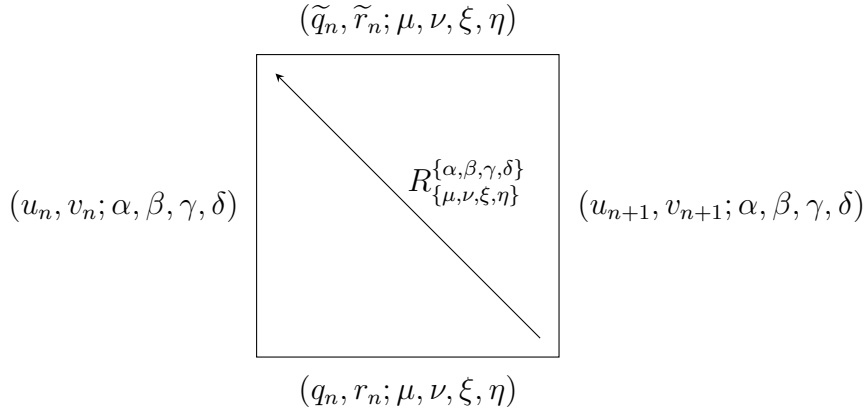
\begin{center}
\begin{tikzpicture}
\draw (0,0) rectangle (4,4);
\draw (2,-0.1)node[below]{$(q_n,r_n; \mu, \nu, \xi, \eta)$}; 
\draw (2,4.1)node[above] {$(\widetilde{q}_n, \widetilde{r}_n; \mu, \nu, \xi, \eta)$} ; 
\draw (-0.1,2)node[left] {$(u_n, v_n; \alpha, \beta, \gamma, \delta)$} ; 
\draw (4.1,2)node[right] {$(u_{n+1},v_{n+1}; \alpha, \beta, \gamma, \delta)$} ; 
\draw[->,>=stealth] (3.75,0.25)--(0.25,3.75);
\draw (1.85,1.85) node[above right] {$R_{\{\mu, \nu, \xi, \eta\}}^{\{\alpha, \beta, \gamma, \delta\}}$}; 
\end{tikzpicture}
\end{center}

The  
map 
\mbox{$R_{\{\mu, \nu, \xi, \eta\}}^{\{\alpha, \beta, \gamma, \delta\}}$} 
described by (\ref{YB_q1})--(\ref{YB_v1}) 
is a parameter-dependent 
Yang--Baxter map 
because its admits 
the Lax representation~\cite{Kouloukas2009,Kouloukas2011}  
given by the 
discrete zero-curvature 
condition (\ref{fd-Lax}). 
If we adopt 
the original definition 
by Veselov and coworkers~\cite{GoVe04,SuVe03,Ve07,BoSu08}, 
the Lax representation for this Yang--Baxter map 
is given by taking the matrix transpose (or matrix inverse) of (\ref{fd-Lax}). 
For the Lax representation to be proper, 
we 
have to 
confirm 
the 
uniqueness 
of 
factorization~\cite{Ve03,GoVe04}   
for 
the product 
of 
three 
Lax matrices~\cite{Kouloukas2009,Kouloukas2011}.  
In our case, 
assume that the relation 
\begin{align}
& {\cal L}  (q', r'; \mu, \nu, \xi, \eta) {\cal L}(u', v'; \alpha, \beta, \gamma, \delta) 
 {\cal L} ({\cal U}', {\cal V}'; a, b, c, d)
\nonumber \\
&=   {\cal L}  (q, r; \mu, \nu, \xi, \eta) {\cal L}(u, v; \alpha, \beta, \gamma, \delta)
 {\cal L} ({\cal U}, {\cal V}; a, b, c, d), 
\label{KP_condition}
\end{align}
holds, 
where the Lax matrices are given by 
(\ref{Ln_YB}), (\ref{Ln_YB2}) and (\ref{Ln_DarBa1}), 
all the parameters are nonzero and 
$\nu/\xi$, $\eta/\mu$, $\beta/\gamma$, $\delta/\alpha$, $b/c$ and $d/a$ 
are 
pairwise distinct. 
Then, 
we should 
have 
\mbox{$q' =q$}, \mbox{$r' =r$}, \mbox{$u' =u$}, \mbox{$v' =v$}, 
\mbox{${\cal U}' ={\cal U}$} and \mbox{${\cal V}' ={\cal V}$}. 
That is, 
the equality (\ref{KP_condition}) holds 
only in the trivial case. 

This 
uniqueness 
of factorization 
can be checked 
as follows. 
With the aid of 
the 
relations 
(\ref{L_qr_inverse}), (\ref{L_uv_inverse}) and 
\begin{equation}
 {\cal L} ({\cal U}, {\cal V}; a, b, c, d) {\cal L} ({\cal U}, {\cal V}; c, d, a, b) = ac I, 
\nonumber
\end{equation}
we can rewrite 
(\ref{KP_condition}) as 
\begin{align}
&  {\cal L}  (q, r; \xi, \eta, \mu, \nu) {\cal L}  (q', r'; \mu, \nu, \xi, \eta) 
\nonumber \\
&= \frac{\mu\xi}{ac\alpha\gamma}{\cal L}(u, v; \alpha, \beta, \gamma, \delta)
 {\cal L} ({\cal U}, {\cal V}; a, b, c, d)  {\cal L} ({\cal U}', {\cal V}'; c, d, a, b) {\cal L}(u', v'; \gamma, \delta, \alpha, \beta). 
\label{KP_condition2}
\end{align} 
Using (\ref{Ln_YB}), 
we can expand 
the left-hand side of (\ref{KP_condition2}) 
as 
\begin{align}
& \left[
\begin{array}{cc}
 \mu\xi I & O \\ 
 O & \mu\xi I \\
\end{array}
\right] 
 - \frac{\mu \nu - \xi \eta}{\mu \zeta^2 + \eta}
 \left[
\begin{array}{c}
 q \\
 -\mu \zeta I \\
\end{array}
\right]
(\mu \nu I - r q)^{-1}  
\left[
\begin{array}{cc}
 \mu r & - \mu \xi \zeta I \\
\end{array}
\right]
\nonumber \\
&
+ \frac{\mu \nu - \xi \eta}{\xi \zeta^2 + \nu}
 \left[
\begin{array}{c}
 \xi q' \\
 -\mu \xi \zeta I \\
\end{array}
\right]
(\xi \eta I - r' q')^{-1}  
\left[
\begin{array}{cc}
 r' & - \xi \zeta I \\
\end{array}
\right]
\nonumber \\
&  - (\mu \nu - \xi \eta) \left( \frac{\mu}{\mu \zeta^2 + \eta} - \frac{\xi}{\xi \zeta^2 + \nu} \right)
 \left[
\begin{array}{c}
 q \\
 -\mu \zeta I \\
\end{array}
\right]
(\mu \nu I - r q)^{-1} r q'
(\xi \eta I - r' q')^{-1}  
\left[
\begin{array}{cc}
 r' & - \xi \zeta I \\
\end{array}
\right]
\nonumber \\
& + (\mu \nu - \xi \eta) \left( \frac{\eta}{\mu \zeta^2 + \eta} - \frac{\nu}{\xi \zeta^2 + \nu} \right) \mu \xi
 \left[
\begin{array}{c}
 q \\
 -\mu \zeta I \\
\end{array}
\right]
(\mu \nu I - r q)^{-1} 
(\xi \eta I - r' q')^{-1}  
\left[
\begin{array}{cc}
 r' & - \xi \zeta I \\
\end{array}
\right]. 
\nonumber 
\end{align}
By 
multiplying  
(\ref{KP_condition2})
both from the left and 
right
by the 
diagonal matrix: 
\[
\left[
\begin{array}{cc}
 I & O \\ 
 O & \frac{1}{\zeta} I \\
\end{array}
\right], 
\]
and 
taking the residue at \mbox{$\zeta^2=-\eta/\mu$} and 
\mbox{$\zeta^2=-\nu/\xi$}, respectively, 
we have \mbox{$r'=r$} and \mbox{$q'=q$}. 
Repeating 
a similar procedure, 
we also have \mbox{$v'=v$}, \mbox{$u'=u$}, \mbox{$ {\cal V}'={\cal V}$} and \mbox{${\cal U}'= {\cal U}$}. 

If the parameters satisfy the conditions 
discussed
in subsection~\ref{subsection2.4}:
\begin{equation}
\gamma = \alpha^\ast, \hspace{5mm} \delta = \beta^\ast,  
\hspace{5mm} \xi = \mu^\ast, \hspace{5mm} \eta=\nu^\ast, 
\nonumber
\end{equation}
the Yang--Baxter 
map described by (\ref{YB_q1})--(\ref{YB_v1}) 
admits 
the Hermitian conjugation reduction; that is, 
the reduction 
\mbox{$r_n= q_n^\dagger$}, 
\mbox{$v_{n+1}=u_{n+1}^\dagger$} 
on the right-hand sides of the equations 
implies the reduction 
\mbox{$\widetilde{r}_n= \widetilde{q}_n^{\,
	\dagger}$}, 
\mbox{$v_{n}=u_{n}^{\, \dagger}$}
on the left-hand sides of the equations. 
Thus, 
we obtain the reduced form of the Yang--Baxter 
map \mbox{$R_{\{\mu, \nu\}}^{\{\alpha, \beta\}}:\ 
\left( q_n, u_{n+1} \right) \mapsto \left( \widetilde{q}_n, u_n \right)$} 
as given by 
\begin{align}
\widetilde{q}_n =& \left(\alpha^\ast \beta^\ast I - u_{n+1} u_{n+1}^\dagger \right)^{-1}\left[ 
-\alpha \beta^\ast (\beta \mu^\ast - \alpha^\ast \nu) q_n + \nu \mu^\ast (\alpha \beta - \alpha^\ast \beta^\ast) u_{n+1}
\right.
\nonumber \\
& 
\left. \mbox{}
- (\alpha \nu - \beta^\ast \mu^\ast) u_{n+1} u_{n+1}^\dagger q_n \right]
\left[ \beta \alpha^\ast (\alpha \nu - \beta^\ast \mu^\ast) I + (\beta \mu^\ast - \alpha^\ast \nu) u_{n+1}^\dagger u_{n+1} 
\right.
\nonumber \\
& \left. \mbox{} - (\alpha \beta - \alpha^\ast \beta^\ast) u_{n+1}^\dagger q_{n} \right]^{-1} 
\left(\alpha^\ast \beta^\ast I - u_{n+1}^\dagger u_{n+1}\right), 
\label{4.8} 
\end{align}
\begin{align}
u_{n} =& \left(\mu \nu I - q_{n} q_{n}^\dagger \right)^{-1} \left[
 \nu \mu^\ast (\alpha \nu^\ast - \beta^\ast \mu) u_{n+1} + \alpha \beta^\ast (\mu \nu -\mu^\ast \nu^\ast) q_n 
- (\alpha \nu - \beta^\ast \mu^\ast) q_n q_n^\dagger u_{n+1} \right]
\nonumber \\
& \times \left[ \mu \nu^\ast (\alpha \nu - \beta^\ast \mu^\ast) I - (\alpha \nu^\ast - \beta^\ast \mu)q_n^\dagger q_n 
- (\mu \nu - \mu^\ast \nu^\ast) q_{n}^\dagger u_{n+1} \right]^{-1} \left(\mu \nu I - q_{n}^\dagger q_{n}\right). 
\label{4.9} 
\end{align}
\begin{center}
\begin{tikzpicture}
\draw (0,0) rectangle (4,4);
\draw (2,-0.1)node[below]{$(q_n; \mu, \nu)$}; 
\draw (2,4.1)node[above] {$(\widetilde{q}_n; \mu, \nu)$} ; 
\draw (-0.1,2)node[left] {$(u_n; \alpha, \beta)$} ; 
\draw (4.1,2)node[right] {$(u_{n+1}; \alpha, \beta)$} ; 
\draw[->,>=stealth] (3.75,0.25)--(0.25,3.75);
\draw (1.85,1.85) node[above right] {$R_{\{\mu, \nu\}}^{\{\alpha, \beta\}}$}; 
\end{tikzpicture}
\end{center}

If we set the values of the parameters as in 
(\ref{concrete_value}), 
i.e., 
\begin{equation}
\alpha=-\gamma=2\mathrm{i}/h, \hspace{5mm} \beta=\delta=1, 
\hspace{5mm} \mu=\xi=1, \hspace{5mm} \nu=-\eta=2\mathrm{i}/\varDelta, 
\hspace{5mm} h, \varDelta \in \mathbb{R}_{\ne 0},  
\nonumber
\end{equation}
and consider the simplest case of the scalar dependent variables, 
we obtain the 
map \mbox{$\left( q_n, u_{n+1} \right) \mapsto \left( \widetilde{q}_n, u_n \right)$} 
in the form: 
\begin{equation} 
\label{4.10} 
\left\{ 
\begin{split}
& \widetilde{q}_n = \frac{
 \left( 1 - \frac{h \varDelta}{4} \right) q_n + \mathrm{i} h u_{n+1}
- \frac{\mathrm{i}h}{2} \left( 1+ \frac{h \varDelta}{4} \right) |u_{n+1}|^2 q_n }
{  \left( 1 + \frac{h \varDelta}{4} \right) + \frac{\mathrm{i}h}{2} \left( 1 - \frac{h \varDelta}{4} \right) |u_{n+1}|^2 
- \frac{h \varDelta}{2} u_{n+1}^\ast q_{n}}, 
\\[1.5mm]
& u_{n} = \frac{ 
 \left( 1 - \frac{h \varDelta}{4} \right) u_{n+1} + \mathrm{i}\varDelta q_n 
- \frac{\mathrm{i}\varDelta}{2}  \left( 1 + \frac{h \varDelta}{4} \right) |q_n|^2 u_{n+1}} 
{ \left( 1 + \frac{h \varDelta}{4} \right) + \frac{\mathrm{i}\varDelta}{2} \left( 1 - \frac{h \varDelta}{4} \right) |q_n|^2 
- \frac{h\varDelta}{2} q_{n}^\ast u_{n+1}}. 
\end{split} 
\right. 
\end{equation}
Note that 
(\ref{4.10}) can be rewritten 
as 
\begin{equation} 
\label{4.11} 
\left\{ 
\begin{split}
& \mathrm{i} \frac{\widetilde{q}_n-q_n}{h} = \frac{
 -\frac{ \mathrm{i} \varDelta}{2} q_n -u_{n+1}
 + |u_{n+1}|^2 q_n + \frac{\mathrm{i} \varDelta}{2}u_{n+1}^\ast q_n^2}
{ \left( 1 + \frac{h \varDelta}{4} \right) + \frac{\mathrm{i}h}{2} \left( 1 - \frac{h \varDelta}{4} \right) |u_{n+1}|^2 
- \frac{h \varDelta}{2} u_{n+1}^\ast q_{n}}, 
\\[1.5mm]
& \mathrm{i} \frac{u_{n}-u_{n+1}}{\varDelta} = \frac{ 
 -\frac{ \mathrm{i} h}{2} u_{n+1} - q_n 
 + |q_n|^2 u_{n+1} +\frac{ \mathrm{i} h}{2} q_n^\ast u_{n+1}^2} 
{ \left( 1 + \frac{h \varDelta}{4} \right) + \frac{\mathrm{i}\varDelta}{2} \left( 1 - \frac{h \varDelta}{4} \right) |q_n|^2 
- \frac{h\varDelta}{2} q_{n}^\ast u_{n+1}}. 
\end{split} 
\right. 
\end{equation}
Clearly, 
(\ref{4.11}) 
can be 
interpreted as 
a fully discrete analog of 
the massive Thirring model in light-cone coordinates (\ref{cMTM}), 
so the Yang--Baxter map  \mbox{$\left( q_n, u_{n+1} \right) \mapsto \left( \widetilde{q}_n, u_n \right)$} 
given by (\ref{4.8}) and (\ref{4.9}) 
indeed 
belongs to the class 
discussed 
in the introduction (this is also true for the matrix-valued case). 
%
%
%

%

%
%
%


\section{Concluding remarks}

In this paper, 
we use 
a discrete Lax matrix representing 
the binary  
B\"acklund--Darboux transformation for the continuous 
derivative nonlinear Schr\"odinger (Chen--Lee--Liu\cite{CLL}) hierarchy as the main building block. 
By solving 
the corresponding 
discrete zero-curvature condition, 
we obtain 
a fully discrete system, 
which provides, with a particular choice of the parameters, 
an integrable discretization 
of the massive Thirring model in light-cone 
(or characteristic) 
coordinates. 
The binary  B\"acklund--Darboux transformation 
for the 
discrete massive Thirring model 
can be constructed in a straightforward 
manner and 
the one-soliton solution is 
presented explicitly. 
Using some symmetry properties of 
the discrete 
Lax matrix, 
we can 
rewrite the 
fully discrete system, 
which includes 
the 
discrete massive Thirring model 
as a particular case, 
as a parameter-dependent 
Yang--Baxter map. 
The 
Yang--Baxter map 
satisfies 
the parameter-dependent version~\cite{Kouloukas2009,Kouloukas2011} of 
the quantum Yang--Baxter equation (\ref{YB_equation}), i.e., 
\begin{equation}
R_{12} (\vt{\alpha}, \vt{\beta}) R_{13} (\vt{\alpha}, \vt{\gamma}) R_{23} (\vt{\beta}, \vt{\gamma}) 
=R_{23} (\vt{\beta}, \vt{\gamma}) R_{13} (\vt{\alpha}, \vt{\gamma}) R_{12} (\vt{\alpha}, \vt{\beta}). 
\nonumber 
\end{equation}
Thus, we obtain a new parameter-dependent Yang--Baxter map 
that admits a natural 
continuous limit for $R_{12}$; the two parameters therein can be understood as 
the lattice parameters. 
Note that this Yang--Baxter map 
can 
also be 
constructed using 
the definition of the unknown functions 
appearing in the discrete Lax matrix 
in terms of the linear 
eigenfunction 
(see (\ref{DBtrans}) and (\ref{Ln_DarBa1}) with (\ref{intermediate_def})). 
It would be interesting to discuss 
how this Yang--Baxter map can 
be characterized within the framework of the matrix modified KP equation 
(see 
\cite{DM2019,DM2020} for the simpler case of 
the matrix KP equation). 


\end{document}